\documentclass{vldb}

\usepackage{graphicx}
\usepackage{balance}  % for  \balance command ON LAST PAGE  (only there!)
\usepackage{subcaption}
\usepackage{hyperref}
\usepackage{listings, multicol}
\usepackage{enumerate}
\usepackage{capt-of}
\usepackage{enumitem}
\usepackage[11pt]{moresize}
\usepackage{xcolor}
\vldbTitle{An IDEA: An Ingestion Framework for Data Enrichment in AsterixDB}
\vldbAuthors{Xikui Wang, Michael J. Carey}
\vldbDOI{https://doi.org/10.14778/3342263.3342628}
\vldbVolume{12}
\vldbNumber{11}
\vldbYear{2019}

\newcommand{\etal}{\textit{et al}.~}

\begin{document}

% ****************** TITLE ****************************************

\title{An IDEA:  An \textbf{\ttlit{I}}ngestion Framework for \textbf{\ttlit{D}}ata \textbf{\ttlit{E}}nrichment\\ in \textbf{\ttlit{A}}sterixDB (Extended Version)}

\numberofauthors{2}

\author{
\alignauthor
Xikui Wang\\
       \affaddr{University of California Irvine}\\
       \email{xikuiw@ics.uci.edu}
% 2nd. author
\alignauthor
Michael J. Carey\\
    \affaddr{University of California Irvine}\\
    \email{mjcarey@ics.uci.edu}
}

\maketitle

\begin{abstract}
Big Data today is being generated at an unprecedented rate from various sources such as sensors, applications, and devices, and it often needs to be enriched based on other reference information to support complex analytical queries. Depending on the use case, the enrichment operations can be compiled code, declarative queries, or machine learning models with different complexities.
For enrichments that will be frequently used in the future,
it can be advantageous to push their computation into the ingestion pipeline so that they can be stored  (and queried) together with the data.
In some cases, the referenced information may change over time, so the ingestion pipeline should be able to adapt to such changes to guarantee the currency and/or correctness of the enrichment results.

In this paper, we present a new data ingestion framework that supports data ingestion at scale, enrichments requiring complex operations, and adaptiveness to reference data changes. We explain how this framework has been built on top of Apache AsterixDB and investigate its performance at scale under various workloads.
\end{abstract}

\section{Introduction}
Traditionally, data to be analyzed has been obtained from one or more operational systems, fed through an Extract, Transform, and Load (ETL) process, and stored in a data warehouse~\cite{Chaudhuri:1997}. In today's Big Data era, the data that people work with is no longer limited to the operational data from a company but also includes social network messages, sensor readings, user click-streams, etc. Data from these sources is generated rapidly and continuously. It becomes increasingly undesirable to stage the data in large batches, process it overnight, and then load it into a data warehouse due to the volume of the incoming stream and the need to analyze current data when making important decisions.

To support the ingestion of continuously generated data and provide near real-time data analysis, streaming engines have been introduced into the Big Data analysis architecture~\cite{botan2009federated, duan2015big, watson2014tutorial}. The incoming data is collected by a streaming engine and then pushed to (or periodically pulled by) a warehouse for later complex data analysis. Adding a streaming engine simplifies the ingestion process for the data warehouse, but it introduces data routing overhead between different systems. To minimize this overhead and simplify the architecture, some systems such as Apache AsterixDB~\cite{grover2015data} have chosen to provide an integrated ingestion facility, data feeds, to enable users to ingest data directly into the system.

The ingested data, such as sensor readings, is often not useful alone in high-level data analysis.
To reveal more valuable insights, it needs to be enriched, e.g, by relating it to reference information and/or applying machine learning models. When the enriched data needs to be queried frequently, its computation is often pushed into the ingestion pipeline and the enriched data is then persisted~\cite{meehan2017streamingETL}. This requires the ingestion framework to have the ability to process incoming data efficiently and to access reference data/machine learning models when needed.

Most streaming engines support incoming data processing, but some only support a limited query syntax~\cite{carbone2015flink, kreps2011kafka}. If a processing task requires existing data from a warehouse, most streaming engines would need to query the warehouse repeatedly, as otherwise they would have to keep a copy of that data locally. Frequent queries to the warehouse would increase the load on the system and incur latency, and maintaining multiple copies of the data would require data migration to keep the data consistent~\cite{meehan2017streamingETL}. Both choices would slow down the enrichment/ingestion pipeline and increase the complexity of building the overall analysis platform.

The data feeds in AsterixDB support data enrichment during ingestion by allowing users to attach user-defined functions (UDFs) to the ingestion pipeline. 
UDFs are a longstanding and commonly available feature in databases. Data enrichment operations can be encapsulated in UDFs and be reused for different use cases. A Java UDF in AsterixDB can be used on a data feed to enrich the incoming data using existing information from resource files. A SQL++ UDF on a data feed can manipulate the incoming data declaratively using a SQL++ query. Currently, however, such a SQL++ function in AsterixDB must be limited to only accessing the content of a given input record, as the execution plan for a SQL++ UDF in general can be stateful. If other data were accessed by a SQL++ function attached to an AsterixDB data feed, its query plan could fail to be evaluated or
 generate intermediate states that neglect changes in referenced data. This limits the expressiveness (usefulness) of data enrichment using SQL++ UDFs in AsterixDB today.

Considering the potential benefits of data enrichment during ingestion, we believe that a data ingestion facility should provide high-performance ingestion for incoming data, a full query syntax support to data enrichment, efficient access to existing data, and adaptiveness to changes in referenced data. With these requirements in mind, we have built a new ingestion framework for Apache AsterixDB. We have improved the scalability and stability of its data feeds and enabled users to attach declarative UDFs on data feeds with support for a full query capability. We have decoupled the ingestion pipeline into layers based on their functionality and life-cycle to improve ingestion efficiency and to allow ingestion pipelines to adapt to data changes dynamically. 

The rest of this paper is organized as follows. We introduce the background information about Apache AsterixDB, its Hyracks runtime engine, and rapid data ingestion in Section~\ref{sec:backgrounds}, and then we discuss how to enrich ingested data at scale in Section~\ref{sec:big_data_enrichment}. In Section~\ref{sec:data_enrichment_for_analyses}, we investigate different strategies for utilizing data enrichment for data analysis and the current limitations of each for providing current and correct data enrichment in time. We explain how we built the new ingestion framework and the techniques used in Section~\ref{sec:building_new}, and we elaborate on the details of the new ingestion framework in Section~\ref{sec:new_ing}. We investigate its performance in Section~\ref{sec:exprs}, review related work in Section~\ref{sec:related_work}, and conclude our work in Section~\ref{sec:conclusions}.

\section{Background}
\label{sec:backgrounds}
In this paper, we use Apache AsterixDB to highlight and address the challenges of enriching incoming data during data ingestion. Here we provide a brief introduction to Apache AsterixDB, its runtime engine Hyracks, and data ingestion.

\subsection{Apache AsterixDB}
\label{sec:asterixdb}
Apache AsterixDB~\cite{alsubaiee2014asterixdb} is an open source Big Data Management System (BDMS). It provides distributed data management for large-scale, semi-structured data. It aims to reduce the need for gluing together multiple systems for Big Data analysis. AsterixDB uses SQL++~\cite{sql++don, ong2014sql++} (a SQL-inspired query language for semi-structured data) for user queries and    
\textit{the AsterixDB Data Model} (ADM) to manage the stored data. ADM is a superset of JSON and supports complex objects with nesting and collections. 
Before storing data in AsterixDB, a user can create a \textit{Datatype}, which describes known aspects of the data being stored, and a \textit{Dataset}, which is a collection of records of a datatype. AsterixDB allows a user to specify a datatype as ``open'' which makes it a minimal, extensible description of the stored data. 
As shown in the example in Figure \ref{ddl:asterixdb_sample}, we can create an open datatype named ``TweetType'' with only two required attributes: ``id'' and ``text''. Tweets containing a variety of additional attributes can be stored and queried in this dataset as well.

\begin{figure}[h]
 \small
\begin{lstlisting}[
           language=SQL,
           basicstyle=\ttfamily,
           showstringspaces=false,
           morekeywords={TYPE, DATASET, CREATE, FEED, WITH},
           commentstyle=\color{gray}
        ]
  CREATE TYPE TweetType AS OPEN {
      id : int64,
      text: string
  };
  CREATE DATASET Tweets(TweetType) 
  PRIMARY KEY id;

\end{lstlisting}
\caption{DDL statements for storing tweets}
\label{ddl:asterixdb_sample}
\end{figure}

\subsection{Hyracks}
\label{sec:hyracks}
\begin{figure}[h]
    \centering
    \includegraphics[width=0.40\textwidth]{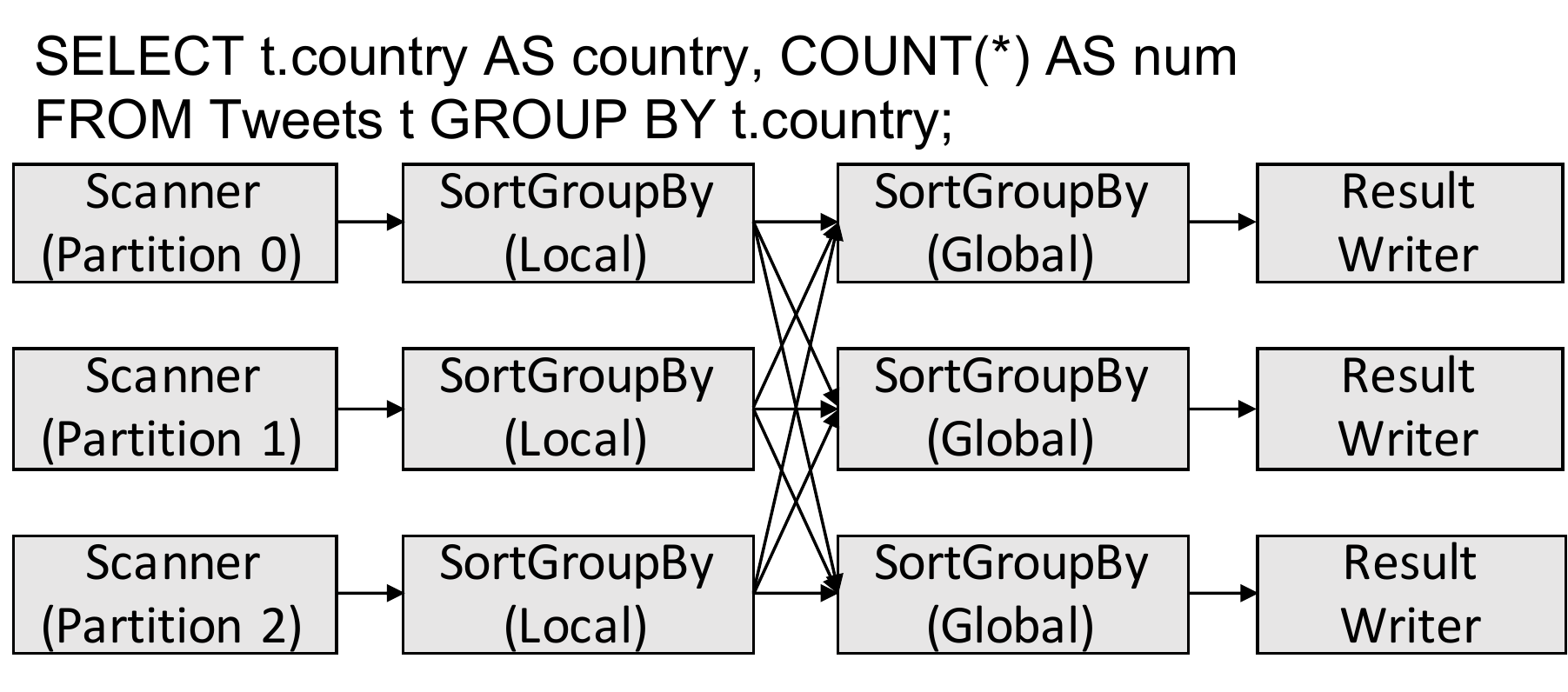}
    \caption{Translating a user query to a Hyracks job}
    \label{fig:hyracks_sample}
\end{figure}

Hyracks~\cite{borkar2011hyracks} is a partitioned parallel computation platform that provides runtime execution support for AsterixDB. Queries from users are compiled into Hyracks jobs. A ``job'' is a unit of work that can be executed on Hyracks. A ``job specification'' describes how data flows and is processed in a job. It contains a DAG of operators, which describe computational operations, and connectors, which express data routing strategies. Data in a  runtime Hyracks job flows in frames containing multiple objects. 
An operator reads an incoming data frame, processes the objects in it, and pushes the processed data frame to another connected operator through a connector. AsterixDB uses jobs to evaluate user queries. A query submitted to AsterixDB is parsed and optimized into a query plan and then compiled as a job specification to run on the  Hyracks platform. Figure~\ref{fig:hyracks_sample} shows an example of how a user query can be represented as a Hyracks job.

\subsection{Data Ingestion}
\label{sec:data_ingestion}
In many contemporary data analysis use cases, data no longer stays on storage devices to be batched into a database system later. Instead, it enters the system rapidly and continuously. The traditional bulk loading technique cannot be applied due to the active nature of the incoming data. Repeatedly issuing insert statements would be impractical because of its low efficiency. 

To untangle database systems and users from the handling of rapidly incoming data, one popular solution is to couple a streaming engine with a database system and use the streaming engine to handle the data~\cite{meehan2017streamingETL}. 
For example, one can set up a Kafka instance to ingest data from external sources, then create a program~\cite{kafka_mongo_diy} or use the \textit{kafka-mongodb-connector}~\cite{kafka_mongo_conn} to transmit the ingested data to a MongoDB instance for later analysis.
A naive solution to this problem could be to create an external program that obtains/receives these records from external sources and pushes them actively into a dataset using \textit{INSERT} statements like the one in Figure~\ref{ddl:ing_insert}.

\begin{figure}[h]
\small
\begin{lstlisting}[
           language=SQL,
           basicstyle=\ttfamily,
           showstringspaces=false,
           morekeywords={TYPE, DATASET, CREATE, FEED, WITH},
           commentstyle=\color{gray}
        ]    
    INSERT INTO Tweets([
      {"id":0, "text": "Let there be light"}
    ]);
\end{lstlisting}
\caption{Insert data into a dataset}
\label{ddl:ing_insert}
\end{figure}

When data comes in at a rapid rate, it is impractical to issue repetitive insert statements for ingesting data into a dataset even if a user batches multiple records into a single insert statement. 
The processing cost of massive insert statements soon becomes a bottleneck in the system and cannot be scaled to handle faster incoming data. In addition, a robust and efficient facility for moving data from external sources into database systems is a common need for many users with similar data analysis use cases. 

While streaming engines provide scalable and reliable data ingestion for database systems, they simultaneously introduce additional data routing costs. For example, the incoming data has to be persisted in Kafka first, then be pushed to the connected database system. Writing the same piece of data multiple times would cost more resources and also delay the demanding analytical queries awaiting the latest ingested data. In addition, configuring multiple systems and wiring them together could be challenging for data analysts, who usually have little system management experience.

To simplify the process of ingesting data for end users and improve ingestion efficiency for analytical systems, some database systems provide integrated data ingestion facilities for handling rapid incoming data. Apache AsterixDB, for example, provides data feeds that allow users to assemble simple data ingestion pipelines with DDL statements~\cite{grover2015data}. 
A user can use the DDL statements in Figure~\ref{ddl:ing_feed} to create a data feed that receives incoming JSON formatted tweets using a socket server with a specified configuration.
A data feed consists of two components: an adapter, which obtains/receives data from an external data source as raw bytes; and a parser, which translates the ingested bytes into ADM records. Compared with the ``glue" solution using streaming engines, data feeds have no extra data routing overheads, and a user can easily assemble a basic data ingestion pipeline with declarative statements. 

\begin{figure}[h]
\small
\begin{lstlisting}[
           language=SQL,
           basicstyle=\ttfamily,
           showstringspaces=false,
           morekeywords={TYPE, DATASET, CREATE, FEED, WITH},
           commentstyle=\color{gray}
        ]    
  CREATE FEED TweetFeed WITH {
    "type-name" : "TweetType",
    "adapter-name": "socket_adapter",
    "format" : "JSON",
    "sockets": "127.0.0.1:10001",
    "address-type": "IP"
  };
  CONNECT FEED TweetFeed TO DATASET Tweets;
  START FEED TweetFeed;
\end{lstlisting}
\caption{Create a socket feed}
\label{ddl:ing_feed}
\end{figure}

\section{Big Data Enrichment}
\label{sec:big_data_enrichment}

\subsection{Motivation}
\label{sec:udf_motivations}
Due to various restrictions, such as the limited bandwidth of infield sensors and predefined data formats from API providers, data coming from external sources may not contain all the information needed for meaningful data analysis. In these scenarios, one can enrich the ingested data with existing knowledge (reference data) or machine learning models to reveal more useful information. For example, the IP address in a log record can be enriched by referencing IP reputation data to see whether there is known threat activity associated with that IP address~\cite{knapp2015industrial}. 
Similarly, an incoming tweet can be enriched by referencing a sensitive key word list to see if its text is potentially threatening. It can also be processed by data mining techniques to extract sentiment and named entities for later analysis~\cite{doyle2014forecasting, mansmann2014discovering}.
Tweets can be enriched by utilizing linguistic processing, semantic analysis, and sentiment analysis techniques and can be used in societal event forecasting systems~\cite{doyle2014forecasting}. Raw data collected by sensor networks can be enriched to support higher level applications, such as improving training effects in sports~\cite{conroy2010enrichment} and building health-care services for medical institutions~\cite{qanbari2015gatica}.

One way of expressing data enrichment requests is to use User-defined Functions (UDFs). This approach allows users to create functions with queries or programs and to reuse them. It enables users to modularize their data enrichment operations and to easily scale their computations on Big Data with the support of a BDMS, like AsterixDB. Since UDFs are available in most databases, the user model and system design around UDFs can be generalized to similar systems.  In this paper, we use the UDF framework in Apache AsterixDB for data enrichment.

\subsection{UDFs for Data Enrichment}
\label{sec:data_enrichment_udf}
AsterixDB supports both Java  and SQL++ in its UDF framework. One can implement a Java UDF that utilizes the provided API to manipulate an input record, or create a SQL++ UDF to enrich input data using a declarative query. For example, we might want to create a UDF that checks whether a given tweet from the U.S. contains the keyword ``bomb". If so, the UDF should add a new field, ``safety\_check\_flag", to the tweet and set it to ``Red". If not, the UDF should add the same new field and sets it to ``Green". Figure~\ref{udf:java_udf1} shows an example of the Java UDF implementation (Java UDF 1) of such a UDF.

\begin{figure}[h]
\scriptsize
\begin{lstlisting}[
           language=Java,
           basicstyle=\ttfamily,
           showstringspaces=false,
           commentstyle=\color{gray}
        ]
...
public void evaluate(IFunctionHelper functionHelper) 
  throws Exception {
  JRecord inputRecord = 
    (JRecord) functionHelper.getArgument(0);
  JString countryCode = 
    (JString) inputRecord.getValueByName("country");
  JString text = 
    (JString) inputRecord.getValueByName("text");

  safetyCheckFlag.setValue(
    countryCode.getValue().equals("US") && 
    text.getValue().contains("bomb") ? "Red":"Green");
  inputRecord.addField("safety_check_flag",
    safetyCheckFlag);
  functionHelper.setResult(inputRecord);
}
...
\end{lstlisting}
\caption{Java UDF 1 for tweet safety check}
\label{udf:java_udf1}
\end{figure}

Although Java UDFs are powerful tools for enriching incoming data, especially when combined with machine learning models, constructing Java UDFs can be more complicated than writing SQL++ queries when expressing the same data enrichment requirements. In addition, a SQL++ UDF can be updated using an \textit{UPSERT} statement instantly while updating a Java UDF requires a recompilation and redeployment process. Figure~\ref{udf:sqlpp_udf1} shows an equivalent SQL++ UDF that performs the safety check for a given tweet (SQL++ UDF 1).

\begin{figure}[h]
\scriptsize
\begin{lstlisting}[
           language=SQL,
           basicstyle=\ttfamily,
           showstringspaces=false,
           morekeywords={TYPE, DATASET, CREATE, FEED, WITH},
           commentstyle=\color{gray}
        ]
  CREATE FUNCTION USTweetSafetyCheck(tweet) {
    LET safety_check_flag = 
      CASE tweet.country = "US" 
           AND contains(tweet.text, "bomb") 
        WHEN true THEN "Red" ELSE "Green"
      END
    SELECT tweet.*, safety_check_flag
  };
\end{lstlisting}
\caption{SQL++ UDF 1 for tweet safety check}
\label{udf:sqlpp_udf1}
\end{figure}

\subsection{Utilizing Existing Knowledge}
\label{existing_knowledge}
In some use cases, a UDF needs to access existing knowledge, such as machine learning models or relevant stored information, for data enrichment. Both Java and SQL++ UDFs can support utilizing existing knowledge. A Java UDF in AsterixDB can load external files during its initialization. A SQL++ UDF can access reference data stored in datasets. To expand on our tweet safety check example in Section~\ref{sec:data_enrichment_udf}, given a list of countries and their sensitive keywords, suppose we want to flag a tweet from a country if it contains one of the keywords associated with that country. For a Java UDF, we can put the country-to-keywords mappings into a local resource file and load it during the UDF initialization. 
Figure~\ref{udf:java_udf2} shows a snippet of the implementation of this Java UDF (Java UDF 2). 
For a SQL++ UDF, we can store the mappings in a ``SensitiveWords" dataset and use a SQL++ query to enrich the input data. 
Figure~\ref{udf:sqlpp_udf2} shows its SQL++ implementation (SQL++ UDF 2).

\begin{figure*}[h]
\small
\begin{lstlisting}[
           language=Java,
           basicstyle=\ttfamily,
           showstringspaces=false,
           commentstyle=\color{gray}
        ]
    ...
    @Override
    public void initialize(IFunctionHelper functionHelper, String nodeInfo) throws IOException {
    	   ...
       BufferedReader fr = Files.newBufferedReader(Paths.get(keywordListPath));
        fr.lines().forEach(line -> {
            String[] items = line.split("\\|");
            keywordList.putIfAbsent(items[1], new LinkedList<>());
            keywordList.get(items[1]).add(items[2]);
        });
    }
    ...
    public void evaluate(IFunctionHelper functionHelper) throws Exception {
        JRecord inputRecord = (JRecord) functionHelper.getArgument(0);
        JString countryCode = (JString) inputRecord.getValueByName("country");
        JString text = (JString) inputRecord.getValueByName("text");
        List<String> keywords = keywordList.getOrDefault(countryCode.getValue(),
                Collections.emptyList());

        for (String keyword : keywords) {
            safetyCheckFlag.setValue(text.getValue().contains(keyword) ? "Red" : "Green");
        }
        inputRecord.addField("safety_check_flag", safetyCheckFlag);
        functionHelper.setResult(inputRecord);
    }
    ...
\end{lstlisting}
\caption{Java UDF 2 for tweet safety check}
\label{udf:java_udf2}
\end{figure*}

\begin{figure}[h]
\scriptsize
\centering
\begin{lstlisting}[
           language=SQL,
           basicstyle=\ttfamily,
           showstringspaces=false,
           morekeywords={TYPE, DATASET, CREATE, FEED, WITH},
           commentstyle=\color{gray}
        ]
  CREATE FUNCTION tweetSafetyCheck(tweet) {
    LET safety_check_flag = CASE 
      EXISTS(SELECT s FROM SensitiveWords s 
          WHERE tweet.country = s.country AND 
          contains(tweet.text, s.word))
      WHEN true THEN "Red" ELSE "Green"
      END
    SELECT tweet.*, safety_check_flag
  };
\end{lstlisting}
\caption{SQL++ UDF 2 for tweet safety check}
\label{udf:sqlpp_udf2}
\end{figure}

Loading external files in a Java UDF is commonly used for necessary configurations that are infrequently updated. However, when an update occurs, such as new keywords being added for a certain country, the resource files on every node will also need to be updated. A reference dataset used in a SQL++ UDF, in contrast, can easily be updated by \textit{INSERT/UPSERT~\footnote{A \textit{UPSERT} statement in SQL++ inserts an object if there is no another object with the specified key. If not, it replaces the previous object with the new one.}} statements.

\section{Data Enrichment for Analysis}
\label{sec:data_enrichment_for_analyses}
The goal of data enrichment is to allow analysts to use the enriched information in analytical queries. One can enrich ingested data \textbf{lazily}, when constructing the analytical queries, or \textbf{eagerly} at ingestion time and then store the enriched results. Enriching data in analytical queries is good for one-time queries, while enriching and storing enriched results allows faster responses for future analytical queries. In this section, we show examples and discuss the implementation of both options.

\subsection{Option 1 - Enrich during Querying}
\label{sec:option1}
For data enrichment used in one-time analytical queries, one can apply enrichment UDFs directly when querying the data. A UDF in an analytical query can be optimized together with the query to produce an optimized query plan.  For example, to find out how many tweets in each country are marked as ``Red", a sample analytical query using a SQL++ UDF 2  is shown in Figure~\ref{udf:option1}. Since the data enrichment is evaluated together with the analytical query, the query response time can be long in the case of complex enriching UDFs. Also, as the enriched data is not persisted, the same  enrichment needs to be computed multiple times for each incoming analytical query.

\begin{figure}[h]
\scriptsize
\begin{lstlisting}[
           language=SQL,
           basicstyle=\ttfamily,
           showstringspaces=false,
           morekeywords={TYPE, DATASET, CREATE, FEED, WITH},
           commentstyle=\color{gray}
        ]

SELECT tweet.country Country, count(tweet) Num 
FROM Tweets tweet
LET enrichedTweet = tweetSafetyCheck(tweet)[0]
WHERE enrichedTweet.safety_check_flag = "Red"
GROUP BY tweet.country;

\end{lstlisting}
\caption{\small An analytical query using SQL++ UDF 2}
\label{udf:option1}
\end{figure}

\subsection{Option 2 - Enrich during Data Ingestion}
\label{sec:option2}
In common use cases, the enriched data may be used repeatedly in analytical queries at different points in time. In such use cases, enriching data \textbf{lazily} in each analytical query separately can be expensive, as it wastes time evaluating the same UDF on the same data multiple times. In such cases, it can be beneficial to instead persist the enriched data and use it for all future analytical queries with similar needs. To allow faster responses to those queries, data enrichment in such use cases is often completed \textbf{eagerly} during the data ingestion process. Here we discuss three different approaches to enriching data during ingestion using Apache AsterixDB.

\subsubsection{Approach 1 - External Programs}
A naive approach to enrich data during ingestion would be to set up an external program that obtains/receives data from data sources, issues DML statements to enrich the collected data, and then inserts the enriched data into a dataset. A sample insert statement that enriches data using SQL++ UDF 2 and inserts the result into a target dataset ``EnrichedTweets"~\footnote{One can create the ``EnrichedTweets" dataset using a DML statement similar to  Figure~\ref{ddl:asterixdb_sample}.}
is shown in Figure~\ref{ddl:appr1}. However, as discussed in Section~\ref{sec:data_ingestion}, issuing repeated insert statements has significant overheads and would not scale well. 

\begin{figure}[h]
\scriptsize
\begin{lstlisting}[
           language=SQL,
           basicstyle=\ttfamily,
           showstringspaces=false,
           morekeywords={TYPE, DATASET, CREATE, FEED, WITH},
           commentstyle=\color{gray}
        ]
  INSERT INTO EnrichedTweets(
    LET TweetsBatch = ([{"id":0, ...}, 
      {"id":1, ...}, ...])
    SELECT VALUE tweetSafetyCheck(tweet) 
    FROM TweetsBatch tweet
  );
\end{lstlisting}
\caption{Enrich and insert collected tweets}
\label{ddl:appr1}
\end{figure}

\subsubsection{Approach 2 - External Programs w/ Data Feeds}
A user can use the basic data feeds feature introduced in Section~\ref{sec:data_ingestion} to improve ingestion performance. The data can first be ingested into a dataset using data feeds, then enriched and stored in another dataset by applying UDFs. A user could set up an external program that repeatedly issues the DML statement in Figure~\ref{ddl:appr2} to initiate data enrichment for ingested data. Depending on the arrival rate of incoming data, the user may issue a new DML statement as soon as the previous one returns to catch up with the ingestion progress when the arrival rate is high, or wait for a certain period to batch the ingested data when the arrival rate is low. Benefiting from data feeds, this approach consumes the incoming data efficiently, even when the data comes in fast, but a user still needs to set up an external program that constantly initiates the data enrichment. In addition, the data is unnecessarily materialized twice since all information in the tweets is kept in the enriched tweets as well.

\begin{figure}[h]
\scriptsize
\begin{lstlisting}[
           language=SQL,
           basicstyle=\ttfamily,
           showstringspaces=false,
           morekeywords={TYPE, DATASET, CREATE, FEED, WITH},
           commentstyle=\color{gray}
        ]
  INSERT INTO EnrichedTweets(
    SELECT VALUE tweetSafetyCheck(tweet) 
    FROM Tweets tweet WHERE tweet.id NOT IN 
      (SELECT VALUE enrichedTweet.id 
        FROM EnrichedTweets enrichedTweet)
  );
\end{lstlisting}
\caption{Enrich and insert ingested tweets}
\label{ddl:appr2}
\end{figure}

\subsubsection{Approach 3 - Data Feeds w/ UDFs}
In order to avoid the unnecessary materialization of incoming data and make the enriched data available to users as soon as possible, we may attach the data enrichment operation directly to the ingestion pipeline so that the ingested data is enriched before it arrives at storage. Apache AsterixDB allows users to attach certain UDFs to data feeds. As an example, a user could attach SQL++ UDF 1 (in Figure~\ref{udf:sqlpp_udf1}) to a data feed using the DDL statement in Figure~\ref{ddl:appr3}. Incoming tweets are first received by the feed adapter, then parsed by the feed parser, and then enriched by the attached UDF. Finally, they are stored in the connected dataset. A Java UDF, such as the UDF in Figure~\ref{udf:java_udf1}, can also be attached to a data feed.

\begin{figure}[h]
\scriptsize
\begin{lstlisting}[
           language=SQL,
           basicstyle=\ttfamily,
           showstringspaces=false,
           morekeywords={TYPE, DATASET, CREATE, FEED, WITH},
           commentstyle=\color{gray}
        ]
  CONNECT FEED TweetFeed 
  TO DATASET EnrichedTweets 
  APPLY FUNCTION USTweetSafetyCheck;
\end{lstlisting}
\caption{Attach a SQL++ UDF to a data feed}
\label{ddl:appr3}
\end{figure}

\subsection{More Complex Enrichment}
\label{sec:apply_udf2}
\subsubsection{Challenges}
In AsterixDB today, UDF 1 can be attached to a data feed directly, as it only accesses the incoming record and does not create any intermediate states. We call this kind of UDF a \textbf{\textit{stateless}} UDF. UDF 2 is different from UDF 1, as UDF 2 accesses external resources (the ``SensitiveWords'' dataset in the case of SQL++, or the equivalent local resource files in the case of Java) and creates intermediate states (such as in-memory hash tables) used for data enrichment. We call this kind of UDF a \textbf{\textit{stateful}} UDF. 

Attaching a stateful UDF to a data feed can be problematic since in some cases the referenced data can itself be modified during the ingestion process, in which case the intermediate states based on the referenced data need to be refreshed accordingly. 
Also, not all complex and stateful SQL++ UDFs can be applied to a continuously incoming data stream directly.
To illustrate the challenges of applying complex and stateful SQL++ UDFs in the ingestion pipeline, here we discuss three possible computing models for attaching UDF 2 to a data feed.

\subsubsection{Model 1 - Evaluate UDF per Record}
A simple computing model for applying a stateful UDF to a feed is to evaluate the attached UDF against each incoming record separately. An incoming record is received and parsed by the feed adapter and parser first, then enriched and persisted in storage. An equivalent insert statement for enriching and persisting one record is shown in Figure~\ref{ddl:model1}. In this model, each collected datum is treated as a new constant record. The attached UDF evaluates each record separately, and any intermediate states will be refreshed from record to record. This allows the UDF to see data changes during the ingestion process, and it imposes no limitations on the applicable query constructs in attached UDFs. However, evaluating the UDF on a per-record basis may introduce a lot of execution overhead. This model cannot be applied in situations where the data arrives rapidly.

\begin{figure}[h]
\small
\begin{lstlisting}[
           language=SQL,
           basicstyle=\ttfamily,
           showstringspaces=false,
           morekeywords={TYPE, DATASET, CREATE, FEED, WITH},
           commentstyle=\color{gray}
        ]
  INSERT INTO EnrichedTweets( 
  LET tweet = { "id": ... }
  SELECT VALUE tweetSafetyCheck(tweet));
\end{lstlisting}
\caption{Enrich and insert a constant record}
\label{ddl:model1}
\end{figure}

\subsubsection{Model 2 - Evaluate UDF per Batch}
To mitigate the execution overhead, one alternative is to batch the collected incoming records, apply the UDF to the batch, and store the enriched records. An equivalent insert statement for enriching a batch of records was shown in Figure~\ref{ddl:appr1}. The records within one batch are enriched using the same reference data, and reference data changes are captured between batches. A larger batch leads to lower execution overhead but less immediate sensitivity to reference data changes; the converse is also true. A user may choose a balance between ingestion performance and sensitivity to reference data changes by tuning the batch size.

\subsubsection{Model 3 - Stream Datasource}
To further reduce execution overheads, the system could attempt to treat the incoming data stream as an infinite dataset and evaluate the attached UDF as if the stream is a normal dataset. An equivalent insert statement is shown in Figure~\ref{ddl:model3}\footnote{The keyword ``\underline{FEED}" is not an actual supported datasource in SQL++, so one cannot run this DDL statement in the Apache AsterixDB system. Here we use it to conceptually denote a continuous feed datasource.}. 

\begin{figure}[h]
\small
\begin{lstlisting}[
           language=SQL,
           basicstyle=\ttfamily,
           showstringspaces=false,
           morekeywords={TYPE, DATASET, CREATE, FEED, WITH},
           commentstyle=\color{gray},
           emph={FEED}, emphstyle=\underbar
        ]
  INSERT INTO EnrichedTweets(
  SELECT VALUE tweetSafetyCheck(t) 
  FROM FEED Tweets t);
\end{lstlisting}
\caption{Enrich and insert records from a feed}
\label{ddl:model3}
\end{figure}

This model would be more efficient than the previous two, as the attached UDF is initialized once for all incoming data. Any pre-computation for enriching the incoming data occurs only once and is used for all incoming data. Although this model would provide the best ingestion performance since it has the smallest execution overhead,  it cannot be used when the attached UDF is \textit{\textbf{stateful}}. Taking SQL++ UDF 2 as an example, when we attach this UDF to a data feed and use this model to compute it, the evaluation would become a join operation between the ``SensitiveWords" dataset and the never-ending feed data source. When there is a more complicated UDF, such as multi-level join and group-by, the evaluation could create more intermediate states and become even harder to evaluate using this model. Here we list three different scenarios of evaluating UDF 2 using Model 3, depending on the join algorithm and the size of the ``SensitiveWords" dataset.

\begin{enumerate}

\item \textit{Hash Join with a small ``SensitiveWords'' dataset}

The evaluation of a hash join operation consists of two phases: build and probe~\cite{shapiro1986join}. In the build phase, the ``SensitiveWords" dataset would be built into a hash table. In the probe phase, the data coming from the Twitter feed would then use the hash table to find the matching records in the ``SensitiveWords" dataset. 

When the ``SensitiveWords" dataset is small, the created hash table can be kept in memory. This allows incoming data to continuously probe the in-memory hash table for enrichment while the ingestion continues as shown in Figure~\ref{fig:case1}. This appears to be a perfect model for this case, but it cannot incorporate the new changes to the ``SensitiveWords" dataset, as the in-memory hash table would be built once and then used throughout the streaming ingestion process.

\begin{figure}[h!]
    \centering
    \includegraphics[width=0.40\textwidth]{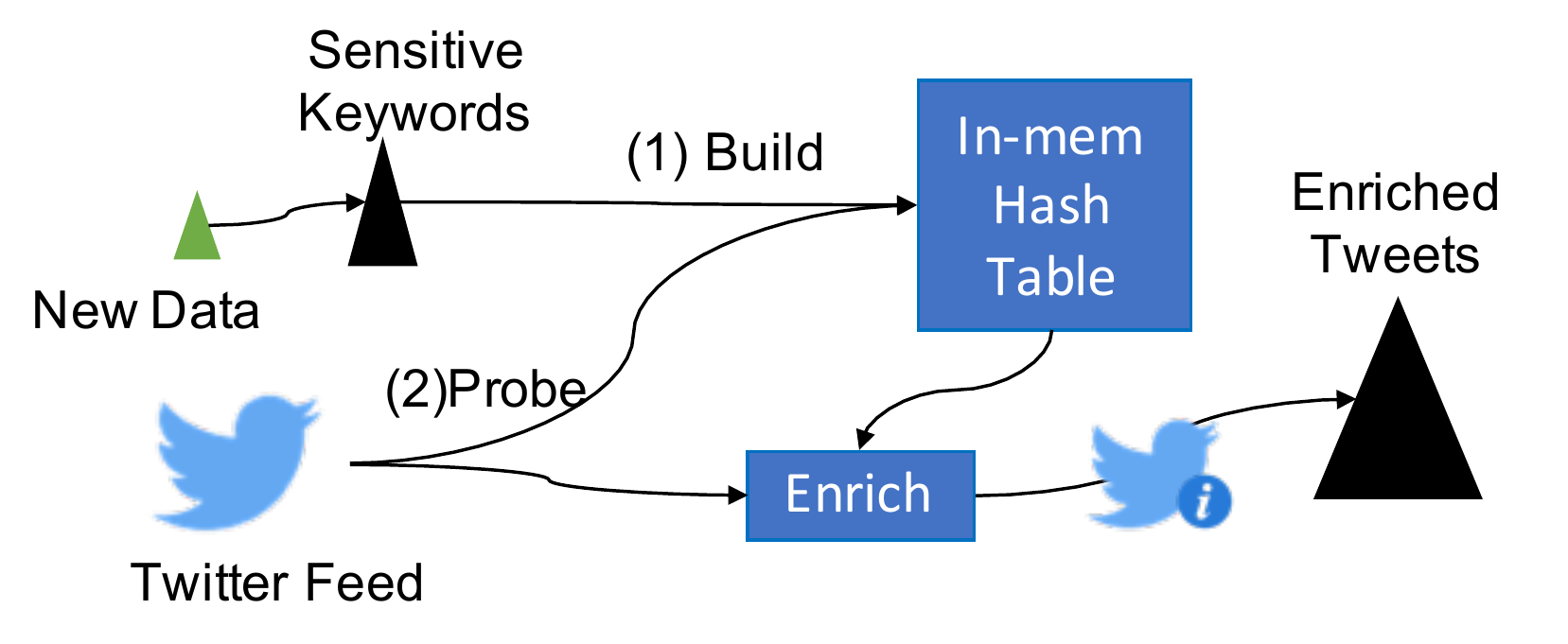}
    \caption{Case 1: Small SensitiveWords dataset}
    \label{fig:case1}
\end{figure}

\item \textit{Hash Join with a big ``SensitiveWords'' dataset}

When the ``SensitiveWords'' dataset is large, part of its data will be spilled to disk for the next round of the join~\cite{shapiro1986join}. This is shown in Figure~\ref{fig:case2}. The hash join algorithm expects to process such spilled data recursively, after reading ``all'' data from Twitter, but of course the tweets will not stop coming. Thus, this model cannot be used in this case.

\begin{figure}[h!]
    \centering
    \includegraphics[width=0.40\textwidth]{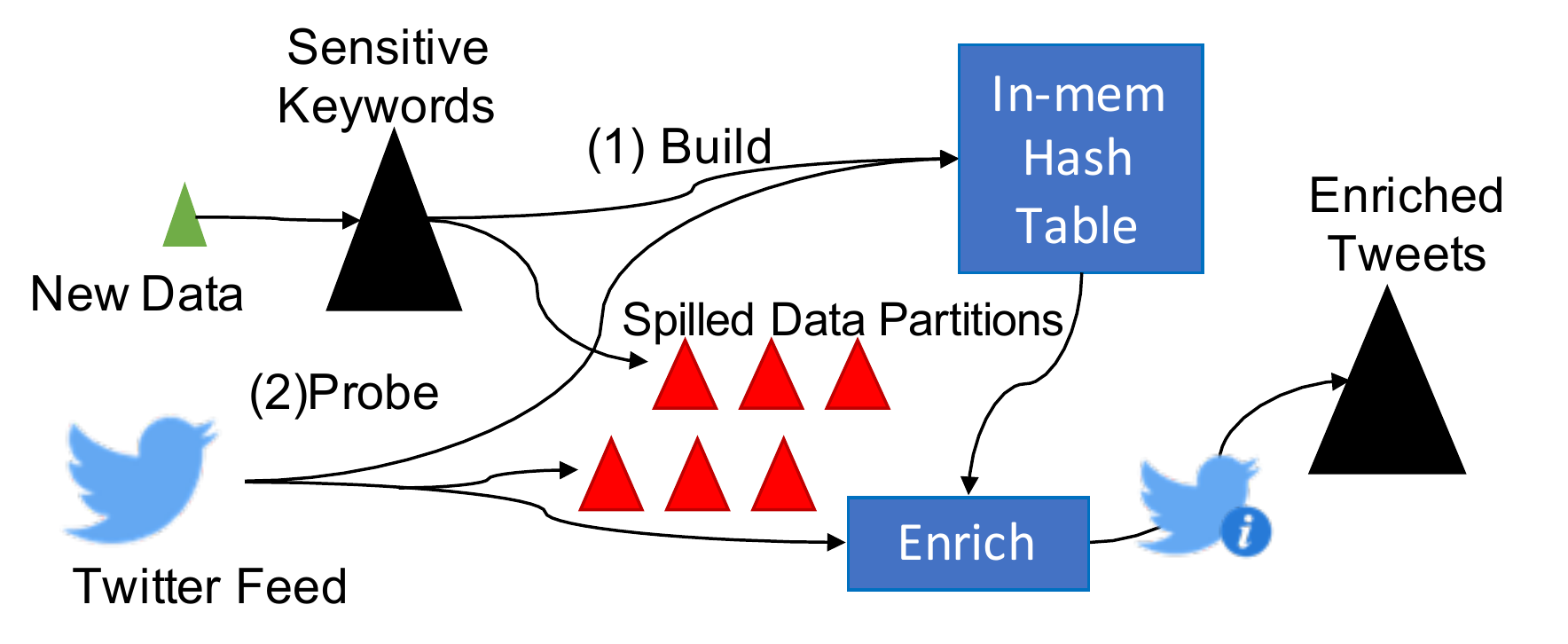}
    \caption{Case 2: Big SensitiveWords dataset}
    \label{fig:case2}
\end{figure}

\item \textit{Index Nested Loops Join}

If there is an index on the ``country'' attribute of the ``SensitiveWords" dataset, the SQL++ query compiler may choose the index nested loop join algorithm to compute the join. In this case, the incoming data can be used to look in the index first, then find the matched records for enrichment, as shown in Figure~\ref{fig:case3}. By choosing this join algorithm manually for this specific join case, one could avoid creating intermediate states during the enrichment operation and thus see the new data changes directly. However, this approach is not applicable to more general use cases where the indexes on referenced datasets may not always exist, and/or where an enrichment UDF contains other operations that create intermediate states. 
As an example, the function in Figure~\ref{udf:udf3} red-flags a tweet if it comes from one of the top 10 countries containing more keywords than others. The top 10 countries list would not be refreshed unless the attached UDF is evaluated again.

\begin{figure}[h!]
    \centering
    \includegraphics[width=0.40\textwidth]{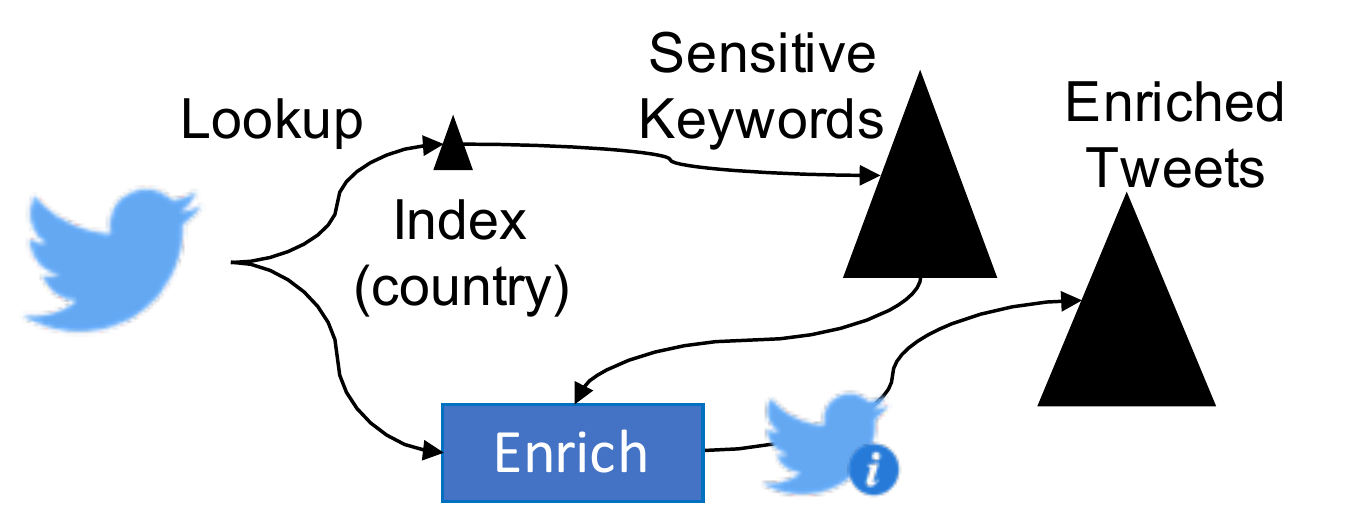}
    \caption{Case 3: Enrich with an available index}
    \label{fig:case3}
\end{figure}

\begin{figure}[h]
\scriptsize
\begin{lstlisting}[
           language=SQL,
           basicstyle=\ttfamily,
           showstringspaces=false,
           morekeywords={TYPE, DATASET, CREATE, FEED, WITH},
           commentstyle=\color{gray}
        ]
  CREATE FUNCTION highRiskTweetCheck(t) {
    LET high_risk_flag = CASE 
      t.country IN (SELECT VALUE s.country
        FROM SensitiveWords s
        GROUP BY s.country
        ORDER BY count(s)
        LIMIT 10)
      WHEN true THEN "Red" ELSE "Green"
    END
    SELECT t.*, high_risk_flag
  };
\end{lstlisting}
\caption{Enrichment UDF with a nested subquery}
\label{udf:udf3}
\end{figure}

\end{enumerate}

In the current Apache AsterixDB release, data feeds actually use this streaming model to evaluate any attached UDFs on an ingestion pipeline, so the attached UDFs are limited to be \textbf{stateless}. In order to support \textbf{stateful} data enrichment UDFs and allow users to use the full power of SQL++ in more complex data enrichment use cases, we need to create a new data ingestion framework that evaluates attached complex \textbf{stateful} UDFs properly.

\section{Framework Building Blocks}
\label{sec:building_new}
As discussed in Section~\ref{sec:apply_udf2}, only models 1 and 2 support complex data enrichment during data ingestion and capture any reference data changes at the same time. We have thus built a new data ingestion framework based on model 2, as it provides flexibility by allowing users to choose the right batch sizes for their use cases. In this section, we describe the design of this new framework and the optimization techniques that we used for improving its performance.

\subsection{Predeployed Jobs}
\label{sec:pjob}
\begin{figure}[ht]
    \centering
    \includegraphics[width=0.40\textwidth]{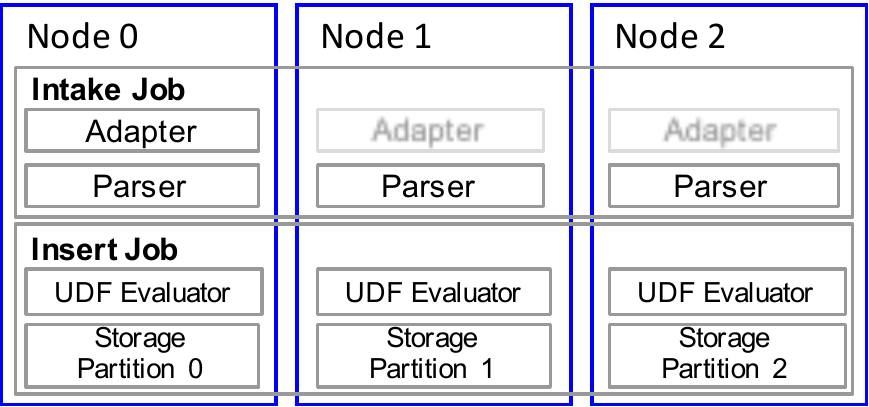}
    \caption{Ingestion pipeline using insert jobs}
    \label{fig:insertjob}
\end{figure}

Following model 2, our new ingestion pipeline consists of two independent Hyracks jobs: an \textit{intake job} and an \textit{insert job}. A sample ingestion pipeline on a three-node cluster is shown in Figure~\ref{fig:insertjob}. The intake job contains the feed adapter
%\footnote{The adapter on each node can be activated separately depending on the use case.} 
and parser, and this job runs continuously throughout the lifetime of the ingestion process. The insert job takes a batch of records from the intake job, enriches them by applying the attached UDFs, and inserts the enriched records into a dataset. It runs repeatedly, being invoked once per batch, during ingestion. In each invocation, the insert job sees the updates to a referenced data record before it is first accessed. Updates after that are picked up by the next invocation\footnote{This follows the record-level consistency model provided in AsterixDB (and most other NoSQL databases).}.

The insert job in Figure~\ref{fig:insertjob} is constructed using the query in Figure~\ref{ddl:appr1}. For every collected batch of records from the intake job, we replace the array of constant records (in TweetsBatch) with the collected batch and execute it. As discussed in Section~\ref{sec:hyracks}, a query in AsterixDB is optimized and compiled into a job specification first, then distributed to the cluster for execution. Since the insert job is executed repeatedly, we utilize \textit{parameterized predeployed jobs} to avoid redundant query compilation and job distribution costs.

\begin{figure}[h!]
    \centering
    \includegraphics[width=0.40\textwidth]{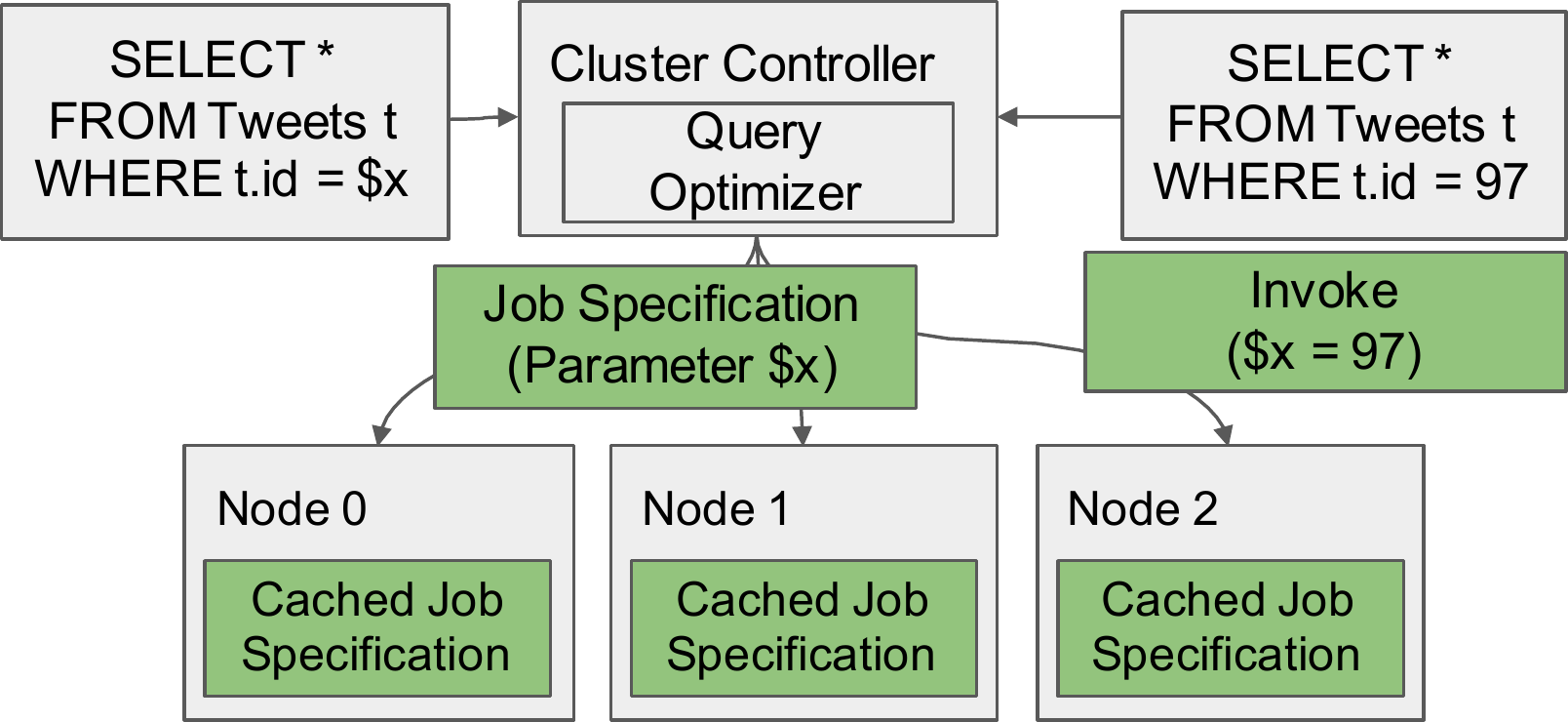}
    \caption{Parameterized predeployed job}
    \label{fig:deployed_job}
\end{figure}

Parameterized predeployed jobs are not unlike prepared queries in relational databases. As shown in Figure~\ref{fig:deployed_job}, a user can choose to predeploy a query with specified parameters. This query is optimized and compiled normally, and then the compiled job specification is predeployed to all nodes in the cluster. This job specification is then cached on the cluster nodes. When a user wants to run this query with a particular parameter, instead of repeating the entire query compilation and distribution process, an invocation message with the new invocation parameter is sent. 
Using this technique, Figure~\ref{fig:insertjob}'s insert job is distributed as a predeployed job in the cluster before the feed starts. When the intake job obtains a new batch of records, it invokes a new insert job with the collected batch as the parameter.

\subsection{Layered Ingestion Pipeline}
\label{sec:layered_pipeline}
\begin{figure}[h]
    \centering
    \includegraphics[width=.45\textwidth]{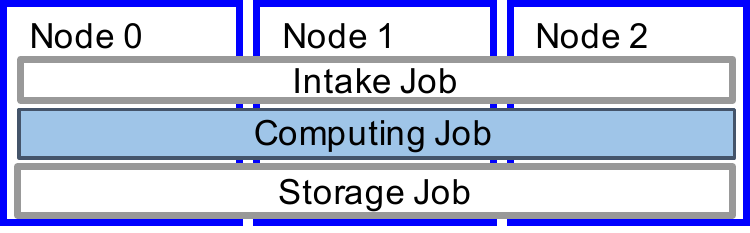}
    \caption{Decoupled ingestion framework}
    \label{fig:decoupled_ing}
\end{figure}

Repeatedly executing the insert job allows any intermediate states created in the UDF evaluation to be refreshed so that any data changes will be used for enriching the incoming data. It should be noted that the evaluation of an insert job, similar to the evaluation of an insert query, will have to wait for the storage log to be flushed to finish properly. Also, since the UDF evaluation and storage operations work sequentially in an insert job, UDF evaluation can be blocked while waiting for the downstream data to be written into storage. To fully utilize the cluster's computing resources and improve overall throughput, we further decompose the insert job into a computing job and a storage job so they can work concurrently. The decoupled ingestion framework is shown in Figure~\ref{fig:decoupled_ing}.

In the decoupled ingestion framework, the intake job handles data from external data sources, the computing job evaluates the attached UDFs, if any, and the storage job writes the enriched data into storage. The intake job and storage job begin to run when the data feed starts, while the computing job in between them is run repeatedly as data batches come in. As in the previous discussion, the computing job is distributed as a predeployed job to reduce execution overhead. Similar to the insert job in Section~\ref{sec:pjob}, an invocation of the computing job will see the updates to a referenced record before it is first accessed by the job.

\subsection{Partition Holders}
In the decoupled ingestion framework, data frames are passed from the intake job to a computing job, and then from the computing job to the storage job. Currently, data exchanges in Hyracks are limited to being within the scope of a job; one job cannot access data frames from another job at runtime. As data exchanges between jobs in the decoupled framework are frequent, we needed to add an efficient mechanism to allow data to be exchanged between jobs.

Considering that the operators in a job each work on data partitions, by aligning the output partitions of one job with the input partitions of the other job, data frames can be shipped from one job to the other efficiently through in-memory structures. For this, we introduce a new type of operator in Hyracks - a partition holder - to enable efficient data exchanges between jobs.

\begin{figure}[h]
    \centering
    \includegraphics[width=.42\textwidth]{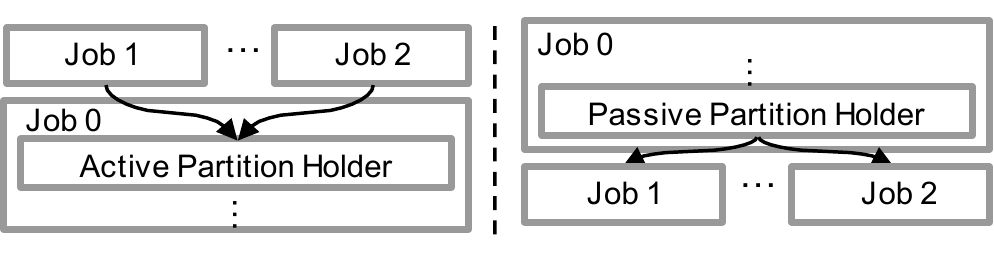}
    \caption{Partition holders}
    \label{fig:partition_holder}
\end{figure}

A partition holder operator ``guards'' a runtime partition by holding the incoming data frames in a queue with a limited size. There are two types of partition holders, active and passive, as shown in Figure~\ref{fig:partition_holder}. An active partition holder follows the default \textbf{push} strategy in Hyracks; it receives data frames from other jobs and pushes them to its downstream operators actively. A passive partition holder implements a \textbf{pull} strategy; it receives data frames from its upstream operators and waits for other jobs to pull them. Each partition holder has a unique ID that is associated with its partition number. When a new partition holder is created, it registers with the local partition holder manager. Jobs sending/receiving data to/from another job can locate the corresponding partition holders through local partition holder managers. In the decoupled ingestion framework, we add a passive partition holder to the tail of the intake job so that the computing jobs can request and receive data in batches. An active partition holder is added to the head of the storage job so that  computing jobs can push the enriched data on to the storage job.

\begin{figure*}[ht!]
    \centering
    \includegraphics[width=0.92\textwidth]{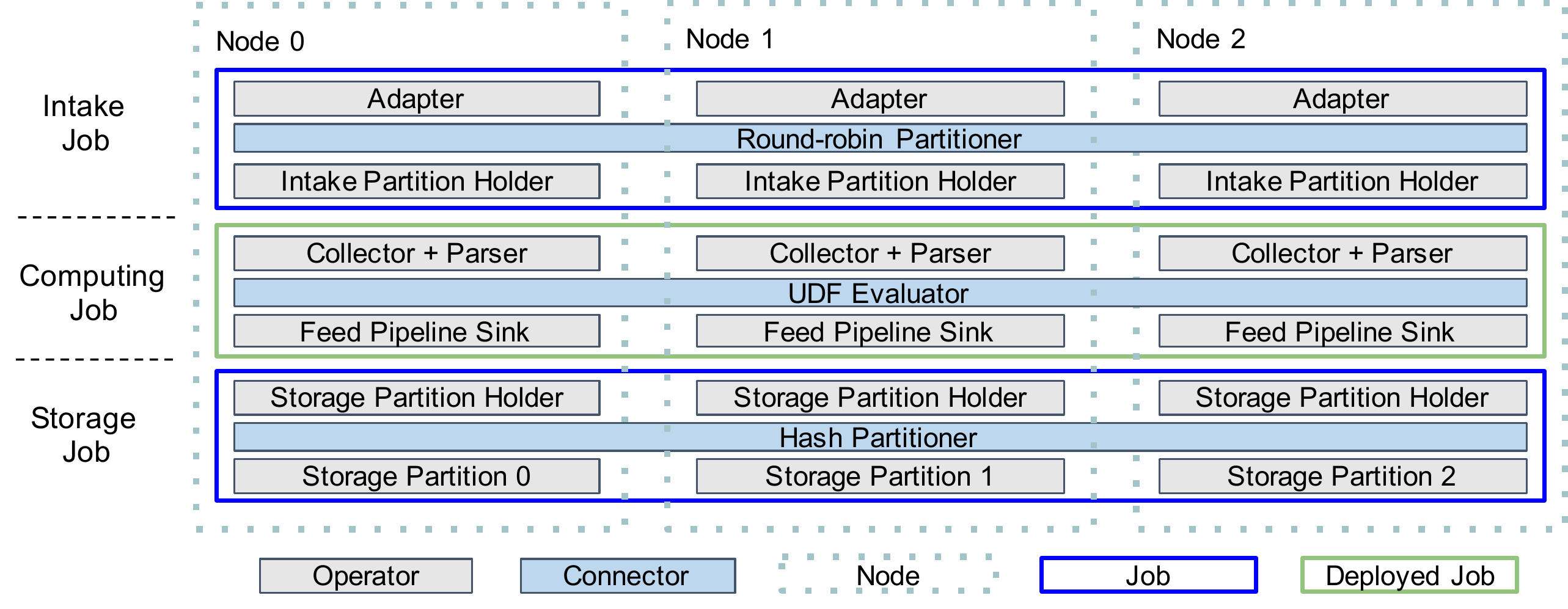}
    \caption{The new ingestion framework}
    \label{fig:architecture}
\end{figure*}

\section{The New Ingestion Framework}
\label{sec:new_ing}
Following the high-level design in Section~\ref{sec:building_new}, we now detail the new ingestion framework in AsterixDB to support data enrichment with reference data updates. We describe how we orchestrate different components in the new ingestion framework in Section~\ref{sec:new_runtime}, and we delve into the lower-level constructs of the framework in Section~\ref{sec:new_architecture}.

\subsection{Ingestion Life Cycle}
\label{sec:new_runtime}
As we discussed earlier, AsterixDB is a parallel data management system that runs on a cluster of commodity machines. In an AsterixDB cluster, one (and only one) node runs the Cluster Controller (CC) that takes in users' queries and translates them into Hyracks jobs. Only the CC can start new jobs, and it keeps track of the progress of the running jobs in case of any failures. All worker nodes in the cluster run a Node Controller (NC) that takes computing tasks from the CC. The CC and NC can coexist on the same node.

In the new ingestion framework, there are two long running jobs, the intake and storage jobs; there is one short lived, but repeatedly invoked, computing job. When there are multiple data feeds running concurrently, each of them is compiled and executed independently. In order to monitor data feed jobs, we created an Active Feed Manager (AFM) on the CC to manage the lifecycle of data feeds. The AFM tracks all active data feeds and helps them to invoke new computing jobs when new data batches arrive. 

When a user submits a start feed request, the CC creates the intake, computing, and storage jobs based on a compiled job specification that is generated from a query template similar to Figure~\ref{ddl:appr1}. The intake and storage job run directly, and the computing job is predeployed into the cluster for later invocations. The AFM maintains the mappings of data feeds to predeployed computing jobs so that it can invoke new computing jobs for each data feed separately.

When an intake job starts, it asks the AFM on the CC to invoke the first computing job and to keep invoking new computing jobs when the previous one finishes. After that, the intake job begins ingesting data from an external data source, adding data records into its queue, and waiting for the computing job to collect the ingested data. The current computing job takes a data batch from the intake job, enriches its records with an attached UDF if any, and then pushes the enriched data batch to the storage job. When this computing job finishes, the AFM on the CC will then start a new computing job to continue the processing.

When the user stops a feed, the intake job first stops taking new data and then adds a special ``EOF" data record into its queue. When a computing job sees this record, it will finish its current execution with the collected data without waiting for a complete batch. The intake job finishes when all ingested data has been consumed. When the intake job finishes, it notifies the AFM to stop invoking new computing jobs for this feed. When the last computing job for the feed finishes, the storage job stops accordingly.

\subsection{New Ingestion Architecture}
\label{sec:new_architecture}
The new ingestion framework consists of the intake, computing, and storage jobs. All jobs run on all nodes in an AsterixDB cluster. As explained in Section~\ref{sec:hyracks}, a Hyracks job contains operators and connectors. In order to demonstrate how data is processed and transported in the new ingestion framework, Figure~\ref{fig:architecture} shows the composition of the framework running on three nodes at the operator and connector level:

\begin{itemize}
\item \textit{The \textbf{intake job}} obtains/receives data from external data sources. The data enters the system through the Adapter.  The Adapter collects data as raw bytes and arranges them into data frames for transportation purposes in the system. A user may choose to activate the Adapter on one or more nodes depending on the expected load. The ingested data frames are then fed through the Round-robin Partitioner to be distributed in a round-robin fashion. Since the attached UDFs can be expensive, distributing the incoming data evenly can help to minimize the overall execution time of the computing job. The partitioned data is forwarded to the Intake Partition Holder, which is implemented as a passive partition holder, which then waits for  computing jobs to pull the data.

\item \textit{The \textbf{computing job}} evaluates the attached UDF to enrich data batches. A computing job starts by collecting a data batch from a local intake partition holder. The obtained data batch is first parsed by the Parser and then fed to the UDF Evaluator for data enrichment. Depending on the attached UDF, the UDF evaluator could be a Java program that runs on each node independently, or it could be a group of operators produced by compiling a complex SQL++ UDF. In either case, local resource files (for Java UDFs) or reference datasets (for SQL++ UDFs) may be accessed, and/or intermediate states might be created as well. After being enriched, the data is pushed to the Feed Pipeline Sink to be forwarded to the storage job.

\item \textit{The \textbf{storage job}} receives enriched data and stores it to disk. The enriched data is first received by the Storage Partition Holder, which is implemented as an active partition holder. A feed pipeline has one Storage Partition Holder on each node, and the Storage Partition Holder receives the enriched data from all local partitions of the collocated computing job. The Storage Partition Holder pushes the received enriched data actively to the connected Hash Partitioner. The Hash Partitioner partitions the enriched data records by their primary keys so they can be stored in the appropriate Storage Partitions.
\end{itemize} 

\section{Experiments}
\label{sec:exprs}
In this section, we present a set of experiments that we have conducted to evaluate the new ingestion framework. We compared the basic ingestion performance of the new ingestion framework with that of the existing Apache AsterixDB ingestion framework. We examined the data enrichment performance of the new ingestion framework using various Java and SQL++ UDFs.  Finally, we investigated the speed-up and scale-out performance of the new ingestion framework for more complex data enrichment workloads. Our experiments were conducted on a cluster which is connected with a Gigabit Ethernet switch. Each node had a Dual-Core AMD Opteron Processor 2212 2.0GHz, 8GB of RAM, and a 900GB hard disk.

\subsection{Basic Data Ingestion}
When ingesting data without an attached UDF, the computing job in the new ingestion framework simply moves data from the intake job to the storage job. By comparing the data ingestion performance of the new ingestion framework to that of the current AsterixDB ingestion framework without UDFs, we can examine the execution overhead introduced by managing and periodically refreshing the computing job in the new ingestion framework. 

\begin{figure}[h]
    \center
    \includegraphics[width=.42\textwidth]{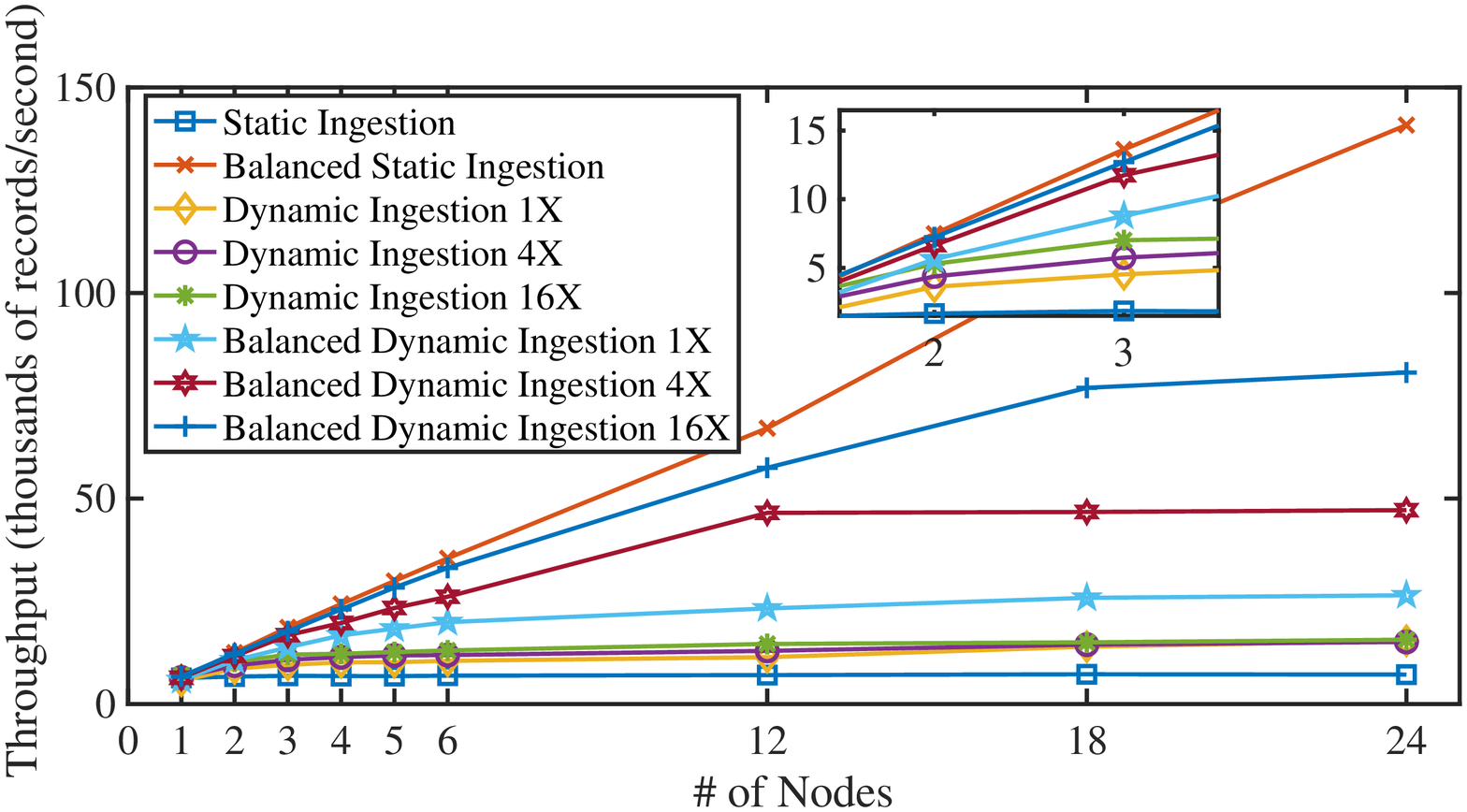}
    \caption{\small 10M tweets ingestion speed-up over 24 nodes}
    \label{fig:ing_perf}
\end{figure}

For this purpose, we compared the throughput of both the current and new frameworks for continuous tweet ingestion. We continuously fed tweets into both ingestion frameworks and measured the throughput of each while consuming 10,000,000 tweets. Each tweet record is around 450 bytes. The results are shown in Figure~\ref{fig:ing_perf}. To make sure a single intake node did not become a bottleneck for the ingestion performance, we also tested a ``balanced version'' of both the current and new ingestion framework by having all nodes in the cluster act as intake nodes. We refer to the experiments on the current framework as ``Static Ingestion", on the new framework as ``Dynamic Ingestion", on the balanced version of the current framework as ``Balanced Static Ingestion", and on the balanced version of the new framework as ``Balanced Dynamic Ingestion''.

To explore how batch size affects the ingestion performance of the new ingestion framework, we experimented with three different batch sizes in Dynamic Ingestion, including 420 records/batch (1X), 1680 records/batch (4X), and 6720 records/batch (16X). Also, we varied the size of the cluster from 1 node to 24 nodes to see how ingestion performance varies with increased cluster sizes.

As we can see in Figure~\ref{fig:ing_perf}, the ingestion performance of Static Ingestion remained the same as the cluster size increased. This is because data intake and parsing are coupled in the current ingestion pipeline. In this case, the ingestion performance was limited by the parsing bottleneck on a single intake node. In contrast, Balanced Static Ingestion kept improving as more nodes participated in parsing and ingesting the data. In the new ingestion framework, data parsing and intake are decoupled, so even for a single intake node (Dynamic Ingestion), the intake performance increased as more nodes participated when the cluster size is small.

Focusing on the results for Dynamic Ingestion, the ingestion performance improved as the batch size increased since there were fewer computing jobs initiated for data enrichment;  different batch sizes' throughputs eventually converged to the same level, as they were limited by the available resources on the single intake node. For the Balanced Dynamic Ingestion, the intake load is split onto all nodes, so its throughput kept growing as nodes were added.

Comparing the performance difference of Balanced Static Ingestion and Balanced Dynamic Ingestion, we can see the execution overhead introduced by repeatedly invoking computing jobs in the new ingestion framework. The execution overhead of invoking computing job increased with the cluster size. As a result, Balanced Dynamic Ingestion had similar throughput as Balanced Static Ingestion when the cluster size is small but started to fall behind as the cluster continued growing. Note that given 24 nodes, the refresh rates (number of computing jobs per second) were 68, 27, and 10 for batch sizes of 1X, 4X, and 16X respectively. We will further explore this with UDFs attached in next section.

\begin{figure*}[htb!]
    \centering
    \includegraphics[width=.98\textwidth]{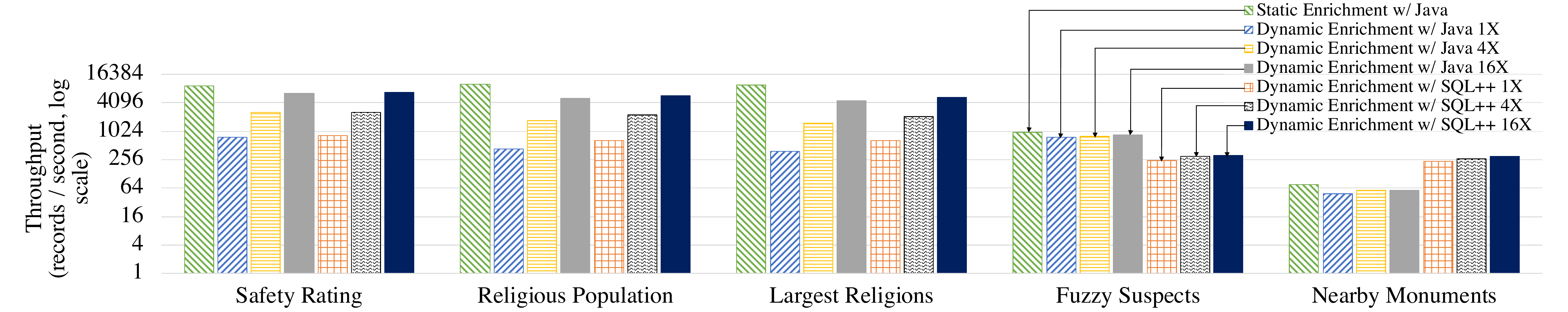}
    \caption{\small 1M tweets Ingestion with UDFs (log scale)}
    \label{fig:udf_type}
\end{figure*}

\subsection{Data Enrichment with UDFs}
\label{sec:udf_expr}
We now turn to the performance of the new ingestion framework in enriching data during data ingestion. We designed four sample use cases where the attached UDFs cover several common operations used in database queries, including join, group-by, order-by, similarity join, and spatial join. The four use cases are as listed below. The reference datasets are SafetyRatings, with 500,000 records and 74 bytes each, ReligiousPopulations, with 500,000 records and 137 bytes each, SuspectsNames, with 5,000 records and 150 bytes each, and MonumentList, with 500,000 records and 94 bytes each.

\begin{enumerate}
    \item \textit{Safety Rating:} Given a list of countries and their corresponding safety ratings, enrich a tweet with a safety rating based on its ``country'' field value. (Hash join. See Appendix~\ref{app:q1})
    \item \textit{Religious Population:} Given the population of each religion in every country, enrich a tweet with the overall religious population based on its ``country'' field value. (Group-by. See Appendix~\ref{app:q2})
    \item \textit{Largest Religions:} Given the population of each religion in every country, enrich a tweet with the three largest religions according to its ``country'' field value. (Order-by. See Appendix~\ref{app:q3})
	\item \textit{Fuzzy Suspects:} Given a list of suspects' names, enrich a tweet with the possible suspects whose name's edit distance to the tweet user's screen name, after removing all special characters, is less than five characters. (Java string processing, Similarity join. See Appendix~\ref{app:q4})
    \item \textit{Nearby Monuments:} Given a list of monuments and their coordinates, enrich a tweet with the monuments within 1.5 degrees of the tweet's location. (Index nested loop spatial join. See Appendix~\ref{app:q5})
\end{enumerate}

All of these enrichment UDFs are stateful, and their evaluations involve the challenges that we discussed in Section~\ref{sec:apply_udf2}. As we have mentioned, the current ingestion pipeline of AsterixDB doesn't support such stateful SQL++ UDFs on its ingestion pipeline. Java UDFs attached on the current AsterixDB ingestion pipeline can only handle reference data without updates. For comparison purpose, we experimented with Java UDFs in current AsterixDB and denote the results as ``Static Enrichment w/ Java".

The new ingestion framework supports both Java and SQL++ UDFs and reference data with updates. We tested both Java and SQL++ UDFs and varied the batch sizes from 420 records/batch to 1680 records/batch to 6720 records/batch to see how batching in the new ingestion framework affects performance. We denote the Java cases as ``Dynamic Enrichment w/ Java 1X'', ``Dynamic Enrichment w/ Java 4X'', and ``Dynamic Enrichment w/ Java 16X " and the SQL++ cases as ``Dynamic Enrichment w/ SQL++ 1X'', ``Dynamic Enrichment w/ SQL++ 4X'', and ``Dynamic Enrichment w/ SQL++ 16X'' respectively.

\begin{figure}[t]
    \centering
    \includegraphics[width=0.45\textwidth]{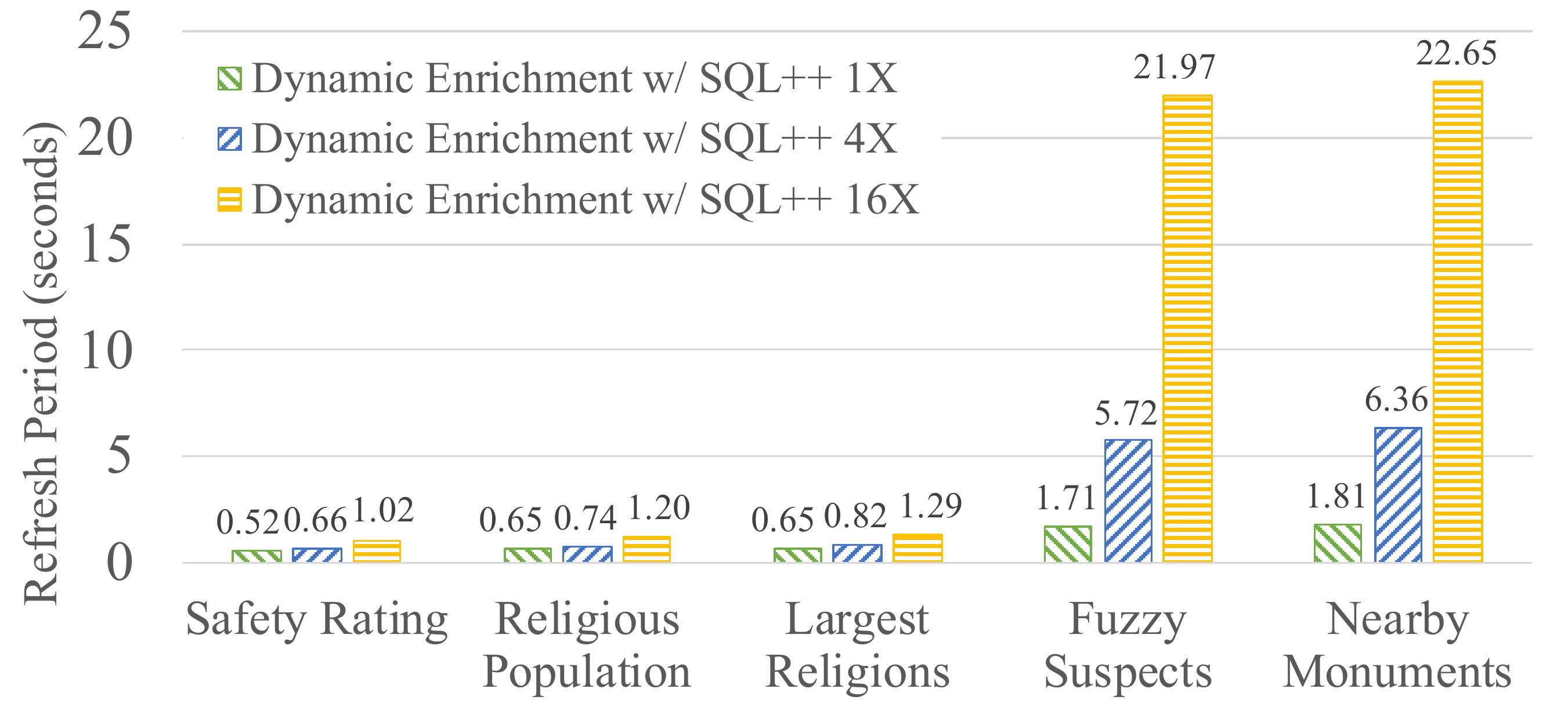}
    \caption{Refresh periods under different batch sizes}
    \label{fig:refresh_period}
\end{figure}

In order to measure the performances of both Static Enrichment and Dynamic Enrichment, we deployed the system on a 6-node cluster and fed tweets to the ingestion pipeline for data enrichment continuously. We measured the throughput (records / second) of the system spent while enriching 1,000,000 tweets, as shown in Figure~\ref{fig:udf_type}  (in log scale). The refresh periods (i.e., execution time per batch) of Dynamic Enrichment w/ SQL++ are shown in Figure~\ref{fig:refresh_period}.

In most of the use cases, except for Nearby Monuments (to be discussed shortly), Static Enrichment offered higher throughput than Dynamic Enrichment. This is because Static Enrichment only loaded reference data once and then reused its stale intermediate states throughout the whole ingestion process. Dynamic Enrichment, however, refreshed and reconstructed those intermediate states from batch to batch. This allowed Dynamic Enrichment to capture data changes to reference data but with the overhead of repeatedly invoking computing jobs. For both Java and SQL++ UDFs, the throughput increased with larger batch sizes since they led to fewer computing jobs and a smaller execution overhead. Accordingly, the refresh periods grew, as there were more records to be enriched in larger batches.

For Fuzzy Suspects and Nearby Monuments, the throughput did not improved that much with increased batch size. The reason was that the computation costs of edit distance and spatial join were high and proportional to the cardinality of the incoming data. Job initialization and management overheads were small relative to these costs. Thus, increasing the batch size didn't improve the throughput significantly. In Fuzzy Suspects in particular, the attached SQL++ UDF not only computed edit distance but also invoked a Java UDF for removing special characters. This introduced extra data serialization/deserialization and shuffling cost. In Nearby Monuments, we created an R-Tree index for the monuments' location in the reference dataset. The use of the index allowed the SQL++ UDF to outperform the Java UDF in this case by performing index lookups on partitioned reference data.

\subsection{Data Enrichment with Updates}
During data ingestion and enrichment, as the referenced data is being updated, the update rate may affect the ingestion performance. In order to further investigate this, we conducted an additional experiment. For each of the use cases (Safety Rating, Religious Population, Largest Religions, Fuzzy Suspects, Nearby Monuments), we created a client program that sends reference data updates to AsterixDB through a  data feed. We measured the resulting throughput impact by varying the update rate (records/second) on a 6-node cluster during 100,000 tweets' ingestion and enrichment.

\begin{figure}[t]
    \centering
    \includegraphics[width=0.45\textwidth]{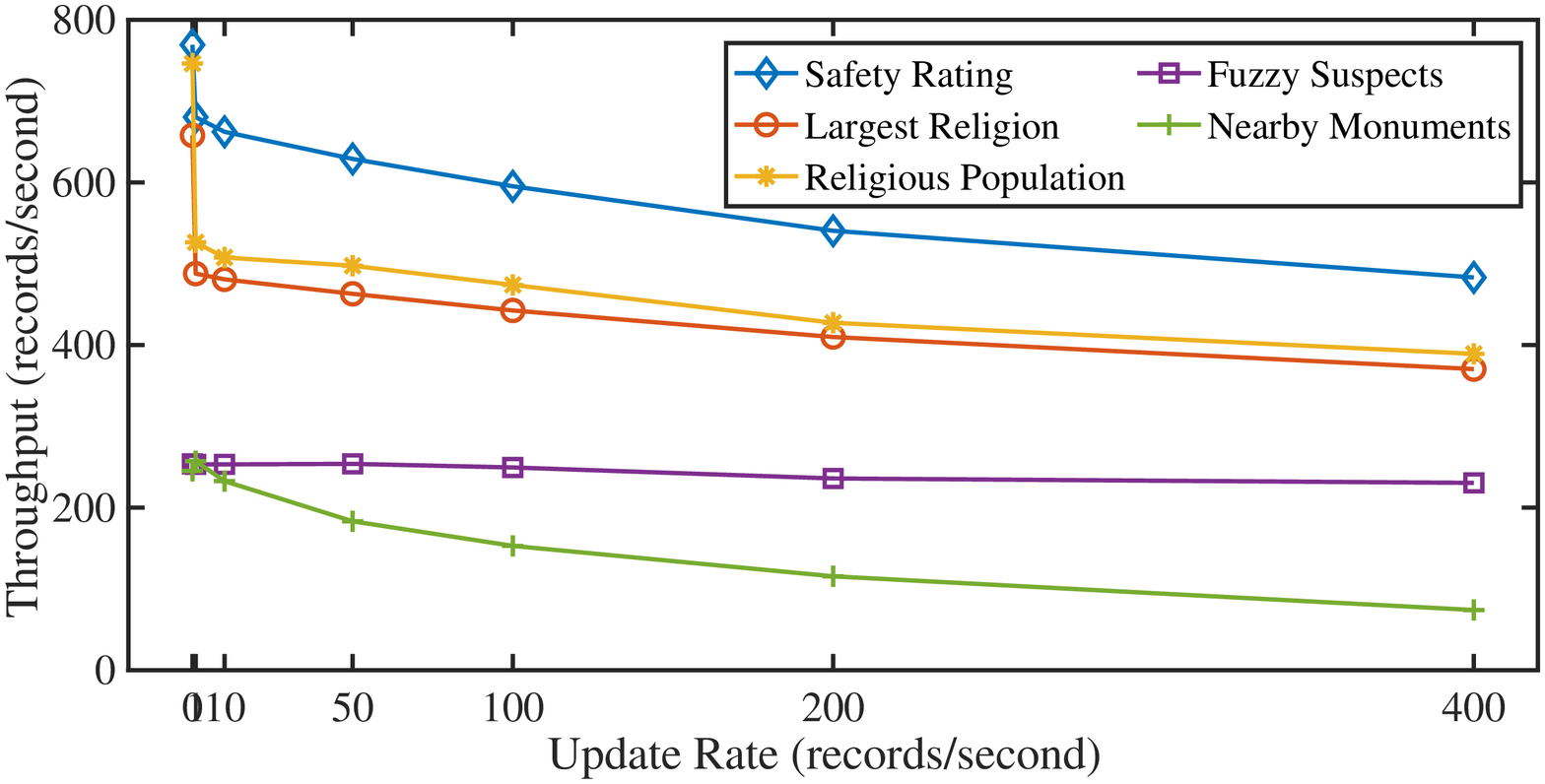}
    \caption{Reference data update}
    \label{fig:ref_update}
\end{figure}

As we can see from Figure~\ref{fig:ref_update}, reference data updates affect the ingestion and enrichment performance differently depending on the cardinality of the referenced dataset and the access method used in data enrichment operations. The throughputs of all cases dropped when the update rate first changed from none to one record per second. This was due to the resulting increased cost in accessing reference data. AsterixDB uses log-structured merge-trees (LSM Trees) in its storage~\cite{alsubaiee2014storage}. Updates to a dataset will activate the in-memory component of its LSM structure and thereby change how the system accesses data even at the low rate of one record per second. This added additional data fetching, locking, and comparison costs to reference data access in computing jobs, which then slowed down the ingestion throughput. Since Fuzzy Suspect had the smallest reference dataset among all, it was the least affected by the updates. When increasing the update rate from there, the throughput then decreases gradually. For Nearby Monuments, the referenced dataset was probed throughout a computing job (index join) for data enrichment whereas in other cases, the referenced dataset was scanned once at the beginning of each computing job (hash join). As a result, Nearby Monuments' performance was less affected at low update rates but started to slip when the update rate was high. The throughput of Nearby Monuments, under the 400 records/second update rate, was only $24\%$ of that without updates. Compared with Safety Rating, the most affected among the other cases, the ratio was $52\%$.
 
\subsection{Scale-out Experiments}
\label{sec:scale_expr}
\subsubsection{Reference Data Scale-out}
\begin{figure}[t]
    \centering
    \includegraphics[width=0.45\textwidth]{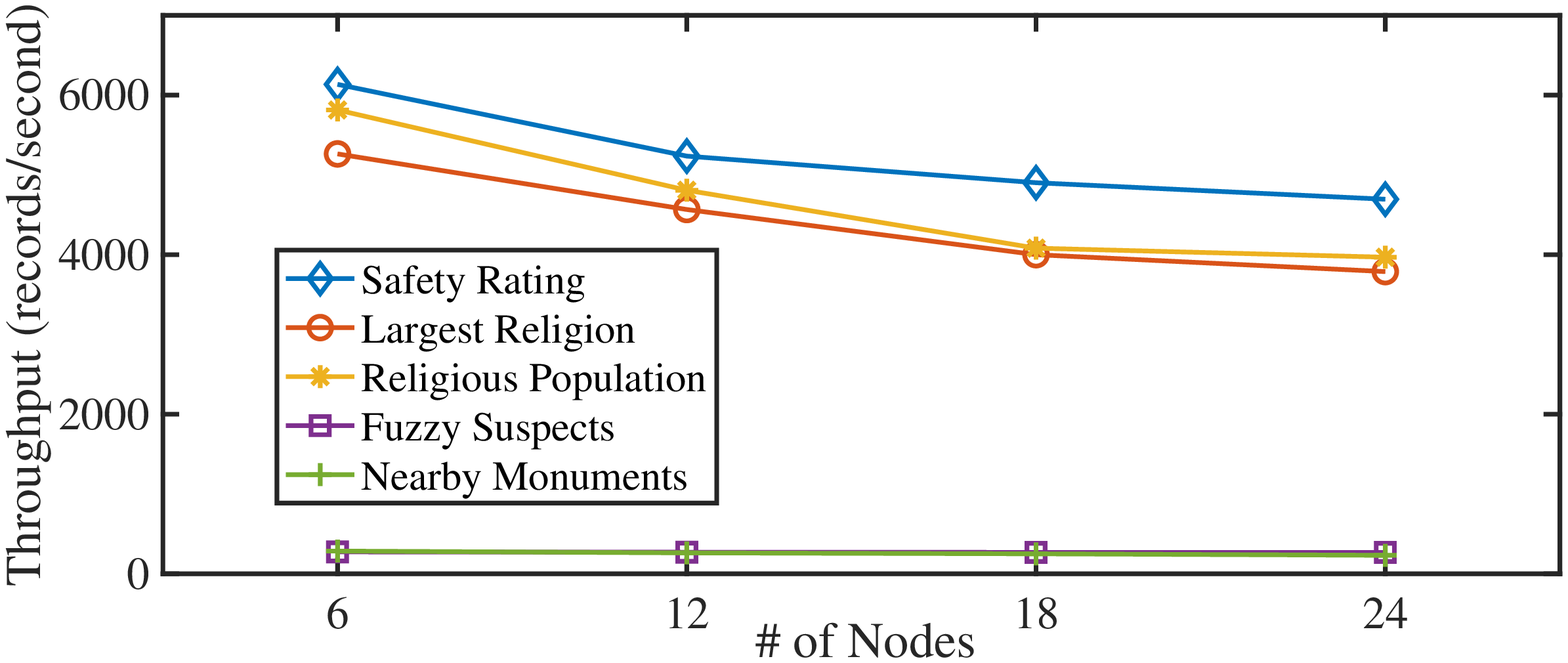}
    \caption{Reference data scale-out}
    \label{fig:ref_scale}
\end{figure}

In the new ingestion framework, since the intermediate states are refreshed repeatedly, the size of the reference data could become a important factor for the data ingestion and enrichment performance. In this section, we explore how the new ingestion framework scales with the size of the reference datasets. We started with the reference datasets in Section~\ref{sec:udf_expr} and increased their sizes to 2X, 3X, and 4X, together with increasing the cluster size to 12 nodes, 18 nodes, and 24 nodes correspondingly. Similarly, we continuously fed tweet data for data enrichment using the same set of SQL++ UDFs and measured the throughput after 1,000,000 tweets with 6720 records/batch. As we can see from the results in Figure~\ref{fig:ref_scale}, the throughput dropped slightly when we increased the size of the cluster due to the increasing execution overhead on a larger cluster. The new ingestion framework scaled well with the reference data size.

\subsubsection{Ingestion Data Scale-out}
In order to further investigate the performance of the new ingestion framework for large scale data enrichment, we designed three new complex data enrichment use cases that add more information to the incoming tweets. The more complex a data enrichment function is, the more performance impact it will have on the whole ingestion framework. We tested the new framework with these  new cases to see how it scales. The additional use cases are listed below. The reference datasets are ReligiousBuildings, with 10,000 records and 205 bytes each, Facilities, with 50,000 records and 142 bytes each, SensitiveNames, with 1,000,000 records and 155 bytes each, AverageIncome, with 50,000 records and 99 bytes each, DistrictArea, with 500 records and 121 bytes each, Residents, with 1,000,000,000 records and 124 bytes each, and AttackEvents, with 5,000 records and 179 bytes each.

\begin{enumerate}[resume]
    \item \textit{Suspicious Names}: Include the number of nearby facilities grouped by their types, the three closest religious buildings within three degrees of the tweet's location, and information about suspicious users who have the same name as the tweet's author. (See Appendix~\ref{app:q6})
    \item \textit{Tweet Context}: Include the average income for the district where the tweet was posted, the number of facilities in this district grouped by their types, and the ethnicity distribution of the residents in this district. (See Appendix~\ref{app:q7})
    \item \textit{Worrisome Tweets}: Include the religion names of the religious buildings within three degrees of the tweet and the number of terrorist attacks in the past two months that were related to that religion. (See Appendix~\ref{app:q8})
\end{enumerate}

\begin{figure}[t]
    \centering
    \includegraphics[width=0.45\textwidth]{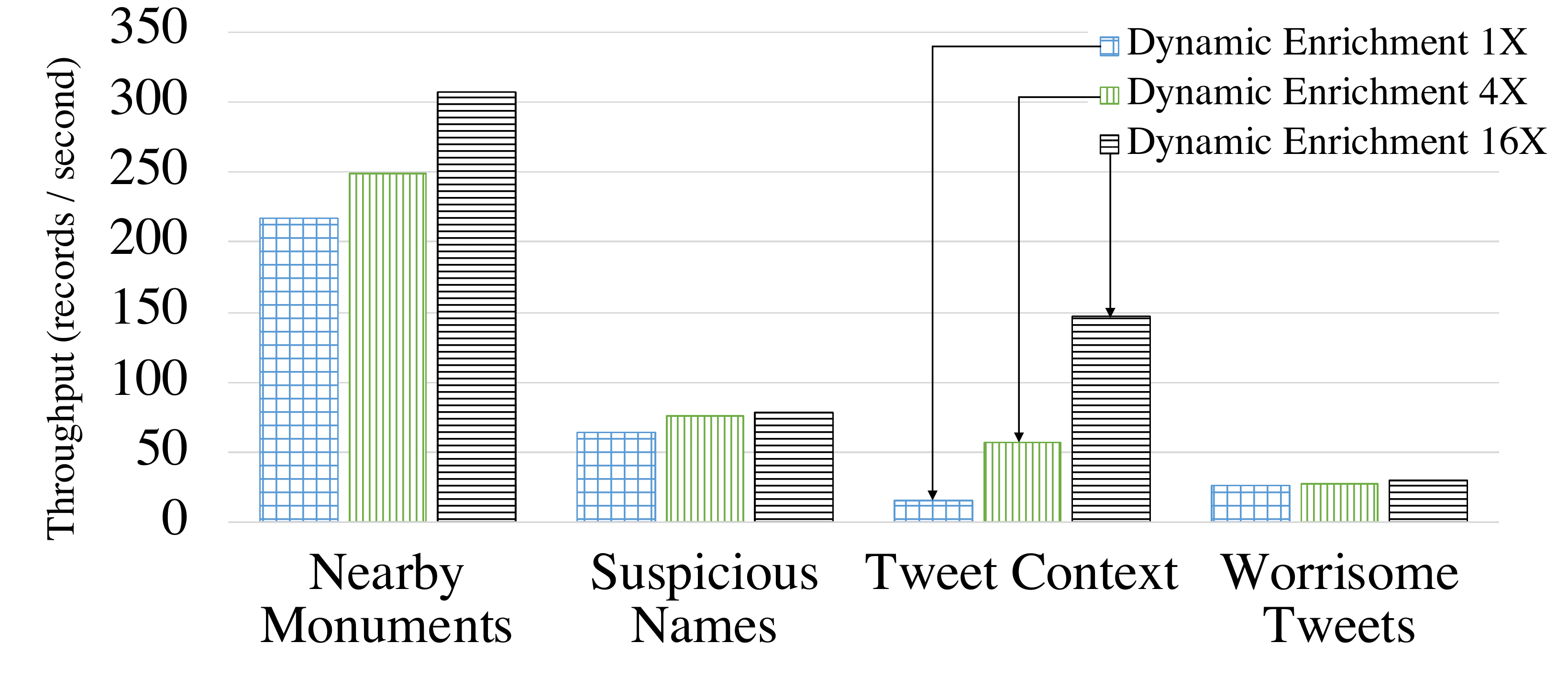}
    \caption{UDF complexity comparison}
    \label{fig:complex_udf}
\end{figure}

\begin{figure*}[ht]
    \center
    \includegraphics[width=.95\textwidth]{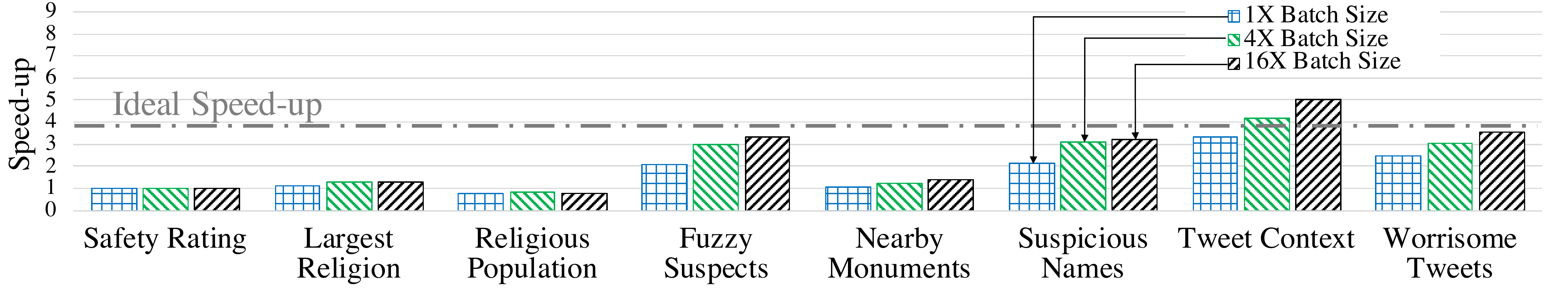}
    \caption{100K tweets ingestion speed-up for 24 vs. 6 Nodes with different batch sizes}
    \label{fig:scale_batchsize}
\end{figure*}

To demonstrate the complexity of these additional use cases, we compared their enrichment performance with that of ``Nearby Monuments'', the most complex UDF from the previous experiment. We measured their throughput on the new ingestion framework for 100,000 tweets enrichment on a 6-node cluster. As shown in Figure~\ref{fig:complex_udf}, the added use cases had different complexities, and different use cases benefited from batch size changes differently. In the Tweet Context use case, there were multiple expensive spatial joins between the referenced datasets before joining with the tweets. Increasing the batch size reduced the computation cost and thus increased the overall ingestion throughput. In the other cases, the tweets mostly joined with the reference datasets sequentially. Thus, increasing the batch sizes offered limited improvements in the Nearby Monuments, Suspicious Names, and Worrisome Tweets use cases.

The performance of scaling out the new framework is determined by the cluster size, batch size, and UDF complexity. Although increasing the number of nodes for computation can reduce the execution time of a computing job, it also introduces additional execution overhead for executing jobs on a larger cluster, so adding more resources may not always improve the overall ingestion time. Given a simple enrichment UDF, a small batch size, and a large cluster, the speed-up performance might be bounded by the batch execution overhead. To  explore the relationship of these three factors, we experimented with the speed-up performance using different batch and cluster sizes.

We let the framework ingest and enrich 100,000 tweets using all seven UDFs.
For each UDF, we measured its throughput on a 6-node cluster and a 24-node cluster separately and computed the resulting speed-up. We repeated this computation for each UDF for three different batch sizes, namely 420 records/batch (1X), 1680 records/batch(4X), and 6720 records/batch (16X), and we show the speed-up of each batch size in Figure~\ref{fig:scale_batchsize}. 

Since the UDFs in the Safety Rating, Religious Population, and Largest Religions use cases were relatively simple and their refresh period is already very low as shown in Figure~\ref{fig:refresh_period}, adding more resources yielded limited improvements to the execution times of their computing jobs. At the same time, their execution overhead grew as the cluster size increased. As a result, their speed-up is relatively poor. In Nearby Monuments, the Index Nested Loop Join didn't benefit from the batch size much as the overall index probing cost is not related to the batch size but mainly to the incoming data cardinality. In contrast, the other UDFs (Fuzzy Suspects, Suspicious Names, Tweet Context, and Worrisome Tweets) each improved with more resources. For Tweet Context in particular, not only were there 4x as many nodes participating in the computation, but the added resources (particularly memory) also allowed the join process to finish earlier. This enabled the system to obtain more than the ideal 4x speed-up. For a given volume of tweets, the bigger the batch size is, the fewer computing job invocations are needed for enriching these tweets, so the speed-up performance is better as the execution overhead increase from the cluster size growth is smaller. 

\begin{figure}[t!]
\centering
\begin{subfigure}[b]{0.45\textwidth}
		\includegraphics[width=\textwidth]{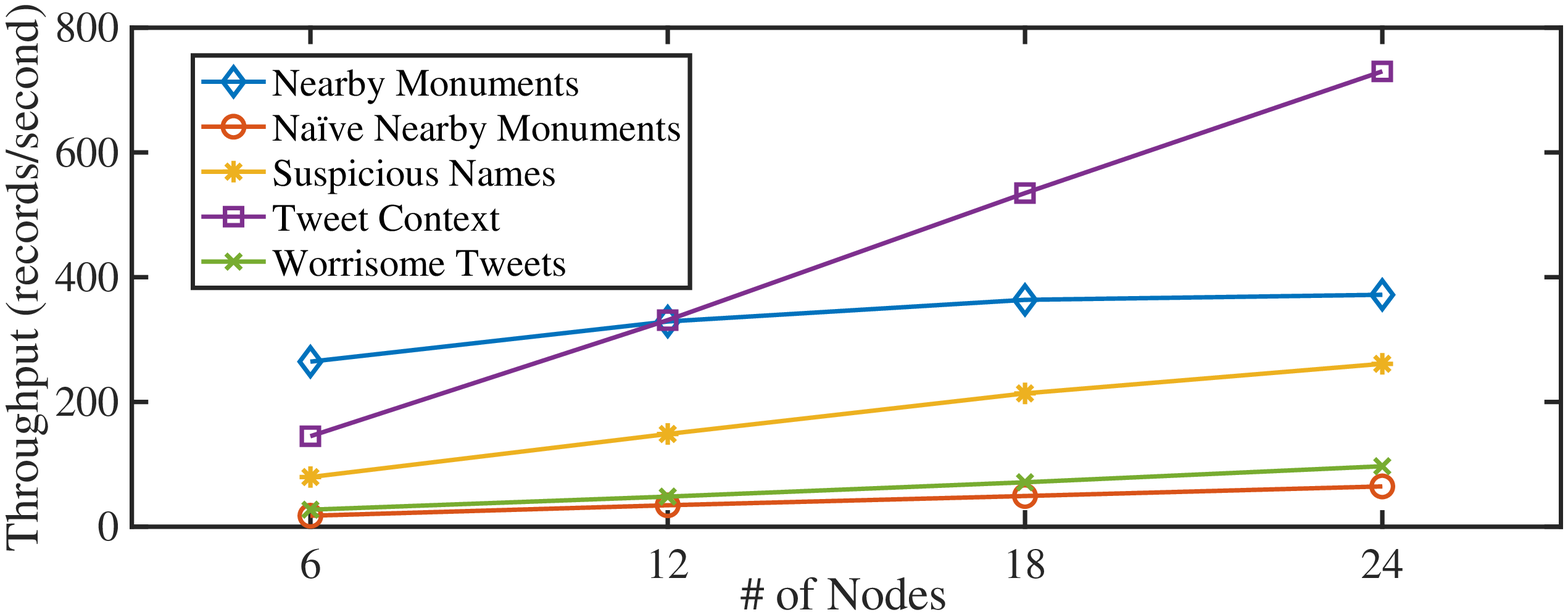}
        \caption{Throughput}
    \end{subfigure}
    \begin{subfigure}[b]{0.45\textwidth}
        \includegraphics[width=\textwidth]{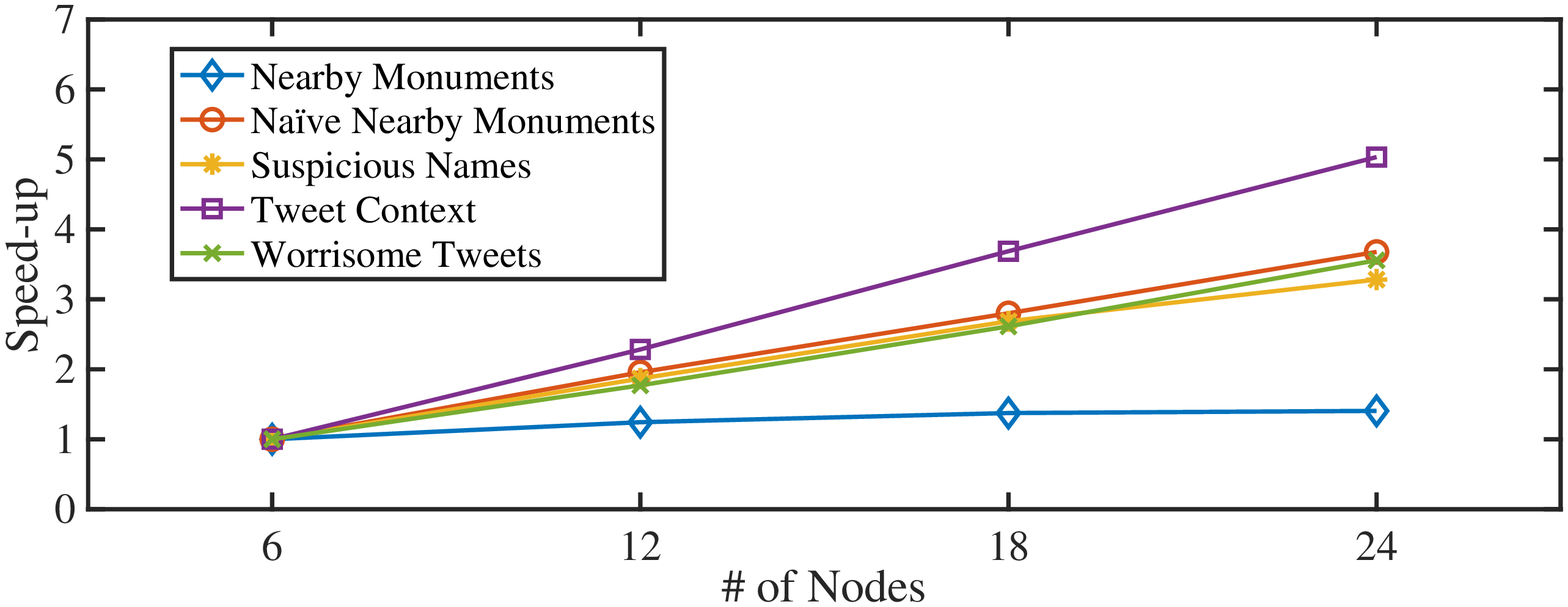}
        \caption{Speed-up}
    \end{subfigure}
    \caption{\small 100K tweets ingestion speed-ups}
    \label{fig:scale_speedup}
\end{figure}

In order to see how ingestion performance improves when adding more resources, we also evaluated the speed-up behavior of the four most complex UDFs (Nearby Monuments, Suspicious Names, Tweet Context, and Worrisome Tweets). To avoid the use of index in Nearby Monuments becoming a performance bottleneck, we used a query hint to add one Naive Nearby Monument use case that enriches the tweets with the same information without using the index.
We fed the new ingestion framework 100,000 tweets and varied the cluster size from 6 nodes up to 24 nodes to see how throughput changes with at batch size of  6720 records/batch. The experimental results are shown in Figure~\ref{fig:scale_speedup}.

As shown,  the ingestion and enrichment performance improved, as more available computing resources were added. The performance gain started to level off when the cluster size kept increasing, as the query execution overhead of a larger cluster started to take away the speed-up benefits for the given reference data sizes. For Nearby Monuments in particular, the Index Nested Loop Join algorithm needed to broadcast the incoming tweets to all nodes to look for intersecting monuments. This limited its speed-up when the cluster size becomes large. In contrast, Naive Nearby Monuments started with a very low throughput which gradually increased as we grew the cluster. The reason was that its reference data monument list was split across more nodes that can then be joined with the incoming tweets concurrently.
 
\section{Related Work}
\label{sec:related_work}
\textbf{Data enrichment} has been widely used in data analysis applications in which the collected data contains limited information and needs to be correlated with existing knowledge to revealing higher-level insights. 
Abel \etal proposed to construct Twitter user profiles by extracting semantics from tweets and relating them with collected news articles~\cite{abel2011semantic}. Moraru \etal introduced a framework for enriching sensor measurements with semantic concepts to generate new features~\cite{moraru2012framework}. In the Big Active Data project~\cite{jacobs2017bad}, notifications delivered to users can be enriched with other existing data in order to provide actionable notifications that are individualized per user; for example, emergency notifications could be enriched with shelter information to help affected users. Our work here is aimed at providing a scalable framework that users can employ to perform such data enrichment operations in the ingestion pipeline so that the enriched data can be used as soon as it is persisted.

\textbf{User-defined functions} have been a long standing feature of database systems~\cite{linnemann1988design,stonebraker1990implementation}. UDFs allow users to register their own functions with the database system for customized data processing and then invoke them in declarative queries. Hellerstein and Stonebraker designed a predicate migration algorithm for moving expensive functions in a query plan to minimize the total cost of a query~\cite{hellerstein1993predicate}. Rheinl{\"a}nder \etal surveyed optimization techniques for optimizing complex dataflows with UDFs~\cite{rheinlander2017dataflow}. In our work, we use the UDF feature as a tool for users to use to express their data enrichment operations. 
Prepared queries are a mechanism that caches compiled plans to improve query performance. The predeployed jobs technique that we employed here for reducing the execution time of computing jobs was inspired by this technique.

\textbf{The traditional ETL process} defines a workflow including data collection, extraction, transformation, cleansing, and loading that is performed for moving data from an operational system into a data warehouse~\cite{Chaudhuri:1997}. Data is extracted from an operational system, cleaned and transformed into a defined schema for analysis, and loaded into a periodically refreshed data warehouse for querying and data analysis. The refreshment process is often executed in an off-line mode with a relatively long period in order to minimize the burden on the operational systems~\cite{vassiliadis2009survey}. Bruckner \etal proposed a near real-time architecture which minimizes the delay of new data being loaded into the data warehouse after being created in the operational system~\cite{bruckner2002striving}. In our work, our focus was building an efficient and succinct framework aimed at ingesting and enriching data at the same time. Note that a user can achieve part of the ETL functionality by constructing appropriate UDFs, we do not consider the new ingestion framework to be a tool for solving general ETL problems. (Similarly, using a complex ETL suite for data enrichment would be overkill.) Our data feeds feature is related to the continuous data loading technique commonly used in near-real-time data warehouses~\cite{jensen2010multidimensional}.

The emerging category of \textbf{hybrid transactional/analyitical processing} (HTAP) aims to serving fast transactional data for analytical requests from large-scale real-time analytics applications. Özcan \etal recently reviewed emerging HTAP solutions and categorized HTAP systems based on different design options~\cite{ozcan2017hybrid}. Some use the same engine to support both OLTP and OLAP requests~\cite{ farber2012sap, raman2013db2}. Other systems choose to couple two separate OLTP and OLAP systems to handle the different workloads separately. Wildfire~\cite{barber2016wildfire}, for example, provides the Wildfire Engine for ingesting fast transactional data, and integrates it with Spark for supporting analytical requests. HTAP systems, and similar data analytics services~\cite{bharadwaj2017creation}, focus on enabling data analytics on recent data. On the other hand, our system looks generally at improving data enrichment performance during the ingestion process so that later analytical queries can be evaluated more efficiently. The techniques that we used in this paper can be adapted to HTAP systems for accelerating their OLAP requests as well.

\textbf{Streaming engines} were introduced to address a need for stream data processing and real-time data analysis. They can handle streaming data sources and provide stream data processing on-the-fly. Many streaming engines also allow users to access reference data during processing. Kafka~\cite{kreps2011kafka}  uses ``change data capture" in combination with its Connect API to access reference data in databases. Flink~\cite{carbone2015flink} supports registering external resources as Tables and offers a DataStream API to process the streaming data. Spark Streaming~\cite{zaharia2016spark} uses Discretized Streams to discretize an incoming stream into Resilient Distributed Datasets and allow users to transform the data using normal Spark operations. Since streaming engines are designed for stream processing but not for complex data analysis queries, the processed results are often stored in connected data warehouses~\cite{meehan2017streamingETL}. In this paper, we have focused on data enrichment use cases where the reference data may be frequently accessed and changed, and where the enriched data needs to be stored in a data warehouse for timely data analysis. We sought to minimize the effort from users so they can create a data ingestion pipeline easily, with declarative statements, and apply enrichment UDFs without limitations. To achieve these goals, we have chosen to build a new ingestion framework that supports the full power of SQL++ for data enrichment operations inside AsterixDB . The batch processing model that we chose is commonly used in streaming engines as well.

\section{Conclusions}
\label{sec:conclusions}
In this paper, we have investigated how to enrich incoming data during the data ingestion process. We discussed the challenges in data ingestion, presented possible computing models for evaluating stateful UDFs for data enrichment, and discussed the problems that may occur in different scenarios. We believe that an ingestion pipeline that supports efficient data ingestion and enrichment should be able to capture reference data changes during the ingestion process, maintain intermediate states properly, and support different enrichment operations with a full query language. To achieve these goals, we created a new ingestion framework with multiple optimization techniques. 
Its layered architecture allows the ingestion pipeline to better utilize the cluster resources. Repeatedly executing computing jobs in the framework allows incoming data to be enriched correctly, and predeployed jobs and partition holders improve the execution efficiency of computing jobs.
We implemented the proposed framework in an open-source DBMS - Apache AsterixDB - and conducted a series of experiments to examine its performance with different workloads and various scales. The results showed that the new ingestion framework can indeed be scaled to support a variety of data enrichment workloads involving reference data and/or stateful operations. The techniques and designs illustrated in this paper could also be applied in other systems to accelerate their analytical requests based on enriched data.

\section{Acknowledgments}
We would like to thank Chen Luo and Vassilis J. Tsotras for their feedback on this paper. The work reported in this paper was supported by the Donald Bren Foundation (via a Bren Chair) and the NSF CNS award 1305430.

\bibliographystyle{abbrv}
\bibliography{vldb_sample} 

\clearpage
\begin{appendix}
In all experiments, we used the data type and dataset defined in Figure~\ref{ddl:asterixdb_sample} for ingesting and storing the tweets.

\section{Safety Rating}
\label{app:q1}
In this experiment, we enrich a tweet with the safety rating of the country where the tweet comes from. 

\begin{figure}[!htbp]
\small
\begin{lstlisting}[
           language=SQL,
           basicstyle=\ttfamily,
           showstringspaces=false,
           morekeywords={TYPE, DATASET, CREATE, FEED, WITH},
           commentstyle=\color{gray}
        ]
  CREATE TYPE SafetyRatingType AS open {
    country_code : string,
    safety_rating: string
  };

  CREATE DATASET SafetyRatings(SafetyRatingType) 
    PRIMARY KEY country_code;

  CREATE FUNCTION enrichTweetQ1(t) {
    LET safety_rating = (SELECT VALUE s.safety_rating 
      FROM SafetyRatings s 
      WHERE t.country = s.country_code)
    SELECT t.*, safety_rating
  };
\end{lstlisting}
\caption{Safety rating enrichment}
\end{figure}

\section{Religious Population}
\label{app:q2}
In this experiment, we enrich an incoming tweet with the total religious population in its country.

\begin{figure}[!htbp]
\small
\begin{lstlisting}[
           language=SQL,
           basicstyle=\ttfamily,
           showstringspaces=false,
           morekeywords={TYPE, DATASET, CREATE, FEED, WITH},
           commentstyle=\color{gray}
        ]
  CREATE TYPE ReligiousPopulationType AS open {
      rid : string,
      country_name : string,
      religion_name : string,
      population: int
  };
  CREATE DATASET ReligiousPopulations
    (ReligiousPopulationType) PRIMARY KEY rid;

  CREATE FUNCTION enrichTweetQ2(t) {
      LET religious_population = 
        (SELECT sum(r.population) FROM 
        ReligiousPopulations r 
        WHERE r.country_name = t.country)[0]
      SELECT t.*, religious_population
  };
\end{lstlisting}
\caption{Religious population enrichment}
\end{figure}

\newpage

\section{Largest Religions}
\label{app:q3}
In this experiment, we enrich a tweet with the 3 largest religions in the country that the tweet came from.

\begin{figure}[!htbp]
\small
\begin{lstlisting}[
           language=SQL,
           basicstyle=\ttfamily,
           showstringspaces=false,
           morekeywords={TYPE, DATASET, CREATE, FEED, WITH},
           commentstyle=\color{gray}
        ]
  CREATE TYPE ReligiousPopulationType AS open {
      rid : string,
      country_name : string,
      religion_name : string,
      population: int
  };
  CREATE DATASET ReligiousPopulations
    (ReligiousPopulationType) PRIMARY KEY rid;

  CREATE FUNCTION enrichTweetQ3(t) {
      LET largest_religions = 
        (SELECT VALUE r.religion_name 
        FROM ReligiousPopulations r 
        WHERE r.country_name = t.country 
        ORDER BY r.population LIMIT 3)
      SELECT t.*, largest_religions
  };
\end{lstlisting}
\caption{Largest religions enrichment}
\end{figure}

\newpage
\section{Fuzzy Suspects}
\label{app:q4}
In this experiment, we use a Java UDF to remove the special characters in a Twitter user's screen name and find the related suspects whose name's edit distance to the processed screen name is within five characters.

\begin{figure}[h]
\scriptsize
\begin{lstlisting}[
           language=Java,
           basicstyle=\ttfamily,
           showstringspaces=false,
           commentstyle=\color{gray}
        ]
    ...
    @Override
    public void evaluate(IFunctionHelper functionHelper) throws Exception {
        JString originalName = (JString) functionHelper.getArgument(0);
        String cleanedString = originalName.getValue().replaceAll("[^a-zA-Z]+", "").toLowerCase();
        cleanedName = (JString) functionHelper.getResultObject();
        cleanedName.setValue(cleanedString);
        functionHelper.setResult(cleanedName);
    }
    ...
\end{lstlisting}
\caption{Java UDF for removing special characters}
\label{udf:java_udf1}
\end{figure}

\begin{figure}[!htbp]
\small
\begin{lstlisting}[
           language=SQL,
           basicstyle=\ttfamily,
           showstringspaces=false,
           morekeywords={TYPE, DATASET, CREATE, FEED, WITH},
           commentstyle=\color{gray}
        ]
  CREATE FUNCTION annotateTweetQ4(x) {
    LET related_suspects=(
      SELECT s.sensitiveName, s.religionName
      FROM SensitiveNamesDataset s
      WHERE edit_distance(
      	testlib#removeSpecial(x.user.screen_name), 
      	s.sensitiveName) < 5)
    SELECT x.*, related_suspects
  };
\end{lstlisting}
\caption{Fuzzy Suspects}
\end{figure}

\section{Nearby Monuments}
\label{app:q5}
In this experiment, we enrich an incoming tweet with the monuments that are within 1.5 degree of the tweet's location.

\begin{figure}[!htbp]
\small
\begin{lstlisting}[
           language=SQL,
           basicstyle=\ttfamily,
           showstringspaces=false,
           morekeywords={TYPE, DATASET, CREATE, FEED, WITH},
           commentstyle=\color{gray}
        ]
  CREATE TYPE monumentType AS open {
    monument_id: string,
    monument_location: point
  };
  CREATE DATASET monumentList(monumentType) 
    PRIMARY KEY monument_id;

  CREATE FUNCTION enrichTweetQ4(t) {
    LET nearby_monuments = 
      (SELECT VALUE m.monument_id 
      FROM monumentList m 
      WHERE spatial_intersect(
        m.monument_location,
        create_circle(
          create_point(t.latitude, t.longitude), 
            1.5)))
    SELECT t.*, nearby_monuments
  };
\end{lstlisting}
\caption{Nearby monuments enrichment}
\end{figure}

\clearpage

\section{Suspicious Names}
\label{app:q6}
In this experiment, we enrich a tweet with the number of nearby facilities grouped by their types, three closest religious buildings within three degrees of the tweet's location, and information about suspicious users who have the same name as the tweet's author.
%\begin{figure*}[htb]
\begin{center}
\small
\begin{lstlisting}[
           language=SQL,
           basicstyle=\ttfamily,
           showstringspaces=false,
           morekeywords={TYPE, DATASET, CREATE, FEED, WITH},
           commentstyle=\color{gray}
        ]
  CREATE TYPE ReligiousBuildingType AS open {
      religious_building_id : string,
      religion_name : string,
      building_location : point,
      registered_believer: int
  };
  CREATE DATASET ReligiousBuildings(ReligiousBuildingType) PRIMARY KEY religious_building_id;

  CREATE TYPE FacilityType AS open {
      facility_id: string,
      facility_location: point,
      facility_type: string
  };
  CREATE DATASET Facilities(FacilityType) PRIMARY KEY facility_id;

  CREATE TYPE SuspiciousNamesType AS open {
      suspicious_name_id: string,
      suspicious_name: string,
      religion_name: string,
      threat_level: int
  };
  CREATE DATASET SuspiciousNames(SuspiciousNamesType) PRIMARY KEY suspicious_name_id;

  CREATE FUNCTION enrichTweetQ5(t) {
      LET nearby_facilities = (
        SELECT f.facility_type FacilityType, count(*) AS Cnt 
        FROM Facilities f 
        WHERE spatial_intersect(create_point(t.latitude, t.longitude), 
          create_circle(f.facility_location, 3.0))
        GROUP BY f.facility_type),
      nearby_religious_buildings = (
        SELECT r.religious_building_id religious_building_id, r.religion_name religion_name
        FROM ReligiousBuildings r
        WHERE spatial_intersect(create_point(t.latitude, t.longitude), 
          create_circle(r.building_location, 3.0))
        ORDER BY spatial_distance(create_point(t.latitude, t.longitude), r.building_location) LIMIT 3),
      suspicious_users_info = (
        SELECT s.suspicious_name_id suspect_id, s.religion_name AS religion, s.threat_level AS threat_level 
        FROM SuspiciousNames s 
        WHERE s.suspicious_name = t.user.name)
      SELECT t.*, nearby_facilities, nearby_religious_buildings, suspicious_users_info
  };
\end{lstlisting}
\captionof{figure}{Enrich a tweet with nearby facilities and suspicious user information}
%\end{figure*}
\end{center}

\clearpage

\section{Tweet Context}
\label{app:q7}
In this experiment, we enrich a tweet with the average income of the district where the tweet is posted, the number of facilities in this district grouped by their types, and the ethnicity distribution of the residents in this district based on a resident sampling.

%\begin{figure*}[!htbp]
\begin{center}
\small
\begin{lstlisting}[
           language=SQL,
           basicstyle=\ttfamily,
           showstringspaces=false,
           morekeywords={TYPE, DATASET, CREATE, FEED, WITH},
           commentstyle=\color{gray}
        ]

  CREATE TYPE DistrictAreaType AS open {
      district_area_id : string,
      district_area : rectangle
  };
  CREATE DATASET DistrictAreas(DistrictAreaType) PRIMARY KEY district_area_id;

  CREATE TYPE FacilityType AS open {
      facility_id: string,
      facility_location: point,
      facility_type: string
  };
  CREATE DATASET Facilities(FacilityType) PRIMARY KEY facility_id;

  CREATE TYPE AverageIncomeType AS open {
      district_area_id: string,
      average_income: double
  };
  CREATE DATASET AverageIncomes(AverageIncomeType) PRIMARY KEY district_area_id;

  CREATE TYPE PersonType AS open {
      person_id: string,
      ethnicity: string,
      location: point
  };
  CREATE DATASET Persons(PersonType) PRIMARY KEY person_id;
  
  CREATE FUNCTION enrichTweetQ6(t) {
      LET area_avg_income = (
        SELECT VALUE a.average_income 
        FROM AverageIncomes a, DistrictAreas d1 
        WHERE a.district_area_id = d1.district_area_id
          AND spatial_intersect(create_point(t.latitude, t.longitude), d1.district_area)),
      area_facilities = (
        SELECT f.facility_type, count(*) AS Cnt 
        FROM Facilities f, DistrictAreas d2
        WHERE spatial_intersect(f.facility_location, d2.district_area) 
          AND spatial_intersect(create_point(t.latitude, t.longitude), d2.district_area)
        GROUP BY f.facility_type),
      ethnicity_dist = (
        SELECT ethnicity, count(*) AS EthnicityPopulation 
        FROM Persons p, DistrictAreas d3 
        WHERE spatial_intersect(create_point(t.latitude, t.longitude), d3.district_area) 
          AND  spatial_intersect(p.location, d3.district_area)
        GROUP BY p.ethnicity AS ethnicity)
      SELECT t.*, area_avg_income, area_facilities, ethnicity_dist
  };
\end{lstlisting}
\captionof{figure}{Enrich a tweet with the average income, facility numbers, and ethnicity distribution in the area where the tweet is posted}
%\end{figure*}
\end{center}

\clearpage

\section{Worrisome Tweets}
\label{app:q8}
In this experiment, we enrich a incoming tweet with the names of religions within three degrees of the tweet's location and the number of terrorist attacks in the past two months related to that religion.

%\begin{figure*}[!htbp]
\begin{center}
%\small
\begin{lstlisting}[
           language=SQL,
           basicstyle=\ttfamily,
           showstringspaces=false,
           morekeywords={TYPE, DATASET, CREATE, FEED, WITH},
           commentstyle=\color{gray}
        ]
  CREATE TYPE ReligiousBuildingType AS open {
      religious_building_id : string,
      religion_name : string,
      building_location : point,
      registered_believer: int
  };
  CREATE DATASET ReligiousBuildings(ReligiousBuildingType) PRIMARY KEY religious_building_id;

  CREATE TYPE AttackEventsType AS open {
      attack_record_id: string,
      attack_datetime: datetime,
      attack_location: point,
      related_religion: string
  };
  CREATE DATASET AttackEvents(AttackEventsType) PRIMARY KEY attack_record_id;

  CREATE FUNCTION enrichTweetQ7(t) {
      LET nearby_religious_attacks = (
        SELECT r.religion_name AS religion, count(a.attack_record_id) AS attack_num
        FROM ReligiousBuildings r, AttackEvents a
        WHERE spatial_intersect(create_point(t.latitude, t.longitude), 
          create_circle(r.building_location, 3.0))
          AND t.created_at  < a.attack_datetime + duration("P2M")
          AND t.created_at  > a.attack_datetime
          AND r.religion_name = a.related_religion
        GROUP BY r.religion_name)
      SELECT t.*, nearby_religious_attacks
  };
\end{lstlisting}
\captionof{figure}{Enrich a tweet with nearby religions and the recent terrorist attacks related to them}
%\end{figure*}
\end{center}

\end{appendix}
%APPENDIX is optional.
% ****************** APPENDIX **************************************
% Example of an appendix; typically would start on a new page
%pagebreak

\end{document}